\documentclass[runningheads,a4paper]{llncs}

\usepackage{epsfig}
\usepackage{graphics}
\usepackage{graphicx}
\usepackage{amsmath}
\usepackage{amsfonts}
\usepackage{amssymb}
\usepackage{url}
\usepackage{subfigure}
\usepackage{multicol}
\usepackage{captcont}

\usepackage{xspace}

\newtheorem{finding}{Finding}

\newcommand{\One}{Berlin\xspace}
\newcommand{\Two}{Hamburg\xspace}
\newcommand{\Three}{Munich\xspace}
\newcommand{\Four}{Cologne\xspace}
\newcommand{\Five}{Frankfurt\xspace}
\newcommand {\Six}{Stuttgart\xspace}
\newcommand {\Seven}{Dortmund\xspace}
\newcommand {\Eight}{Bremen\xspace}
\newcommand {\Nine}{Dresden\xspace}
\newcommand {\OneZero}{Hanover\xspace}
\newcommand {\OneOne}{Leipzig\xspace}
\newcommand {\OneTwo}{Nuremberg\xspace}
\newcommand {\OneThree}{Bielefeld\xspace}
\newcommand {\OneFour}{Mannheim\xspace}
\newcommand {\OneFive}{Karlsruhe\xspace}
\newcommand {\OneSix}{M\"{u}nster\xspace}
\newcommand {\OneSeven}{Augsburg\xspace}
\newcommand {\OneEight}{Aachen\xspace}
\newcommand {\OneNine}{Chemnitz\xspace}
\newcommand {\TwoZero}{Braunschweig\xspace}
\newcommand {\TwoOne}{Kiel\xspace}

\begin{document}

\title{Schlauschleimer in Reichsautobahnen:  \\ Slime mould imitates \\ motorway network in Germany}

\titlerunning{Adamatzky A. and Schubert T. Kybernetes 41 (2012) 7/8, 1050--1071. }

\author{Andrew Adamatzky\inst{1} and Theresa Schubert\inst{2}}

\authorrunning{Schlauschleimer in Reichsautobahnen}

\institute{University of the West of England, Bristol, United Kingdom \\ \email{andrew.adamatzky@uwe.ac.uk}
\and
Bauhaus-Universit\"{a}t Weimar, Weimar, Germany \\ \email{theresa.schubert-minski@uni-weimar.de}}

\maketitle

\begin{centering}
Final version is published in \\
\textbf{
Andrew Adamatzky, Theresa Schubert, (2012) "Schlauschleimer in Reichsautobahnen: Slime mould imitates motorway network in Germany", Kybernetes, Vol. 41 Iss: 7/8, pp.1050--1071.}
\end{centering}

\begin{abstract}

The German motorway, or 'autobahn', is characterised by long traditions, 
meticulous state planning, historical misbalance between West and East Germany's 
transport networks, and highest increase in traffic in modern Europe, posing a need for 
expansion and/or restructuring. We attempt to evaluate the structure of autobahns 
using principles of intrinsic optimality of biological networks in experiments with 
slime mould of \emph{Physarum polycephalum}. In laboratory experiments with living slime mould
we represent major urban areas of Germany with sources of nutrients, inoculate the 
slime mould in Berlin, wait till the slime mould colonises all major urban areas and compare 
the statistical structure of protoplasmic networks with existing autobahn network. The straightforward comparative
analysis of the slime mould and autobahn graphs is supported by integral characteristics and indices of the graphs. 
We also study the protoplasmic and autobahn networks in the context of planar proximity graphs. 

\vspace{0.5cm}

\end{abstract}

\section{Introduction}

In contrast to other countries, motorways in Germany are not just means of transportation but a pivotal part of citizens' mentality. The history of the German motorway network dates back to 1926 when the so-called Hafraba association was formed to build a motorway linking Hamburg with Basel via Frankfurt~\cite{rothengatter_2005}. The project was dormant till Hitler's National Socialist Party came to power and made the whole idea of free-flowing vehicular transport networks a revolutionary driving force --- \emph{Reichsautobahnen}, dictatorship  of motorways --- of the miraculous economical and technological progress of pre-war Germany and an instrument for propaganda. After the Second World War the 'Hitler Autobahns' were re-interpreted as means of reconstruction of West German democracy~\cite{zeller_2007}. The 'Green Nazi' environmentally oriented approach towards the construction of autobahns and the aesthetics and design of surrounding landscapes was --- at different stages of history --- propagandised, mythologised, refuted, dismissed~\cite{zeller_2007} but then partially 'resurrected' in 1990s and now actively considered by specialists in the context of transport network integrative development in Germany. 
  
German motorways, or 'autobahns', are a unique road system because it is amongst earliest state planned transport networks; it was planned precisely and meticulously yet possesses a wide range of quality; and, it  has the highest traffic load in Europe. Peculiar features of German motorways are based on long-standing traditions of car ownership, the wide range of roads quality and the absence of speed limits in some parts of the motorways~\cite{brilon_1994}.  Germany occupies central geographical position in Europe. This leads to a dramatic increase of traffic on autobahns, which in turns leads to increase in number of traffic accidents. Expansion of the autobahn network is amongst many possibilities of solving the traffic problem~\cite{garnowski_2011}. However the possibilities for adding new autobahn routes are very limited.  

How would the autobahn network develop from scratch under the current configuration of urban areas in Germany? Are autobahns optimal from primitive living creatures point of view? Does the topology of autobahns satisfy any principle of 
natural foraging behaviour and fault tolerance? Are there any matches between German transport networks and basic planar proximity graphs? All these are --- at least partially --- answered in the present paper by physically imitating 
the autobahns development in laboratory experiments with slime mould \emph{Physarum polycephalum}. 

\emph{Physarum polycephalum} is an acellular slime mould~\cite{stephenson_2000}. It inhabits forests in many parts of the world, and can be found in under logs and decayed tree branches. Its vegetative stage --- the 
\emph{plasmodium} ---  is a single cell, visible by an unaided eye, with myriad of nuclei. The plasmodium 
feeds on a wide range of microorganisms. During its development and foraging behaviour the plasmodium makes blob-like colonies on sources of nutrients. The colonies are connected in a single organism by 
a network of protoplasmic tubes. The network is considered to be optimal~\cite{nakagaki_2000,nakagaki_2001a} in terms of efficiency of nutrients spanning, sensorial inputs and cost-efficient transportation of nutrients and 
metabolites in the plasmodium's body. In his pioneering works Toshiyuki Nakagaki and his colleagues demonstrated
that the plasmodium's foraging behaviour can be interpreted as a computation~\cite{nakagaki_2000,nakagaki_2001a}. In slime mould 'computers'~\cite{PhysarumMachines} data are represented by spatial configurations of attractants and repellents. The computation is implemented during the slime mould's propagation and colonisation of nutrients. The results of the computation are represented by the structure of the plasmodium's protoplasmic network as developed on a data set of nutrients~\cite{nakagaki_2000,nakagaki_2001a,PhysarumMachines}. The problems solved by plasmodium of \emph{P. polycephalum} include shortest path~\cite{nakagaki_2000,nakagaki_2001a}, implementation of storage modification machines~\cite{adamatzky_ppl_2007}, Voronoi diagram~\cite{shirakawa},  Delaunay triangulation~\cite{PhysarumMachines}, logical computing~\cite{tsuda_2004}, and process algebra~\cite{schumann_adamatzky_2009}.

An early evaluation of the road-modelling potential of  \emph{P. polycephalum} in 2007~\cite{adamatzky_UC07} came to no definite conclusion. However, significant progress has been made since that time; 
such as has been reported in our recent papers on approximation of highways systems in the United Kingdom~\cite{adamatzky_jones_2009},
Mexico~\cite{adamatzky_Mexico}, the Netherlands~\cite{adamatzky_Netherlands}, Iberia~\cite{adamatzky_alonso_2010} and Brazil~\cite{adamatzky_brazil}. For all of these countries we found that the network of protoplasmic tubes developed by \emph{P. polycephalum} matches, at least partly, network of human-made transport systems;  though the closeness of fit varies from country to country.  A variable that likely plays an important role in determining the closeness of fit is the constraint that each country's government design policies place on their unique highway transport networks.
This is why we are in the process of collecting data on the development of plasmodium networks in all major countries, in preparation toward undertaking a final comparative analysis. In the present paper we imitate the development of German transport networks with the slime mould. We believe the result possesses a particular value due to the unique history of autobahns and relatively precise planning of the transport system by German governments over last eighty years. 

The paper is structured as follows. We outline our experimental techniques and methods of analysis of the slime mould and autobahn graphs in Sect.~\ref{experimental}. Particulars of foraging behaviour of the slime mould on sources of nutrients, representing major urban areas of Germany, are discussed in Sect.~\ref{colonisation}. Analysis of protoplasmic networks, developed by \emph{P. polycephalum}, and their comparison with German motorway network are undertaken
in Sects.~\ref{physarumgraphs} and \ref{comparison}. The slime mould and autobahn graphs are considered in terms of planar proximity graphs in Sect.~\ref{proximity}. And finally, in Sect.~\ref{disaster}
we study a response of protoplasmic network to imitate large-scale contamination.

\section{Experimental and methods}
\label{experimental}

\begin{figure}[!tbp]
\centering
\includegraphics[width=0.8\textwidth]{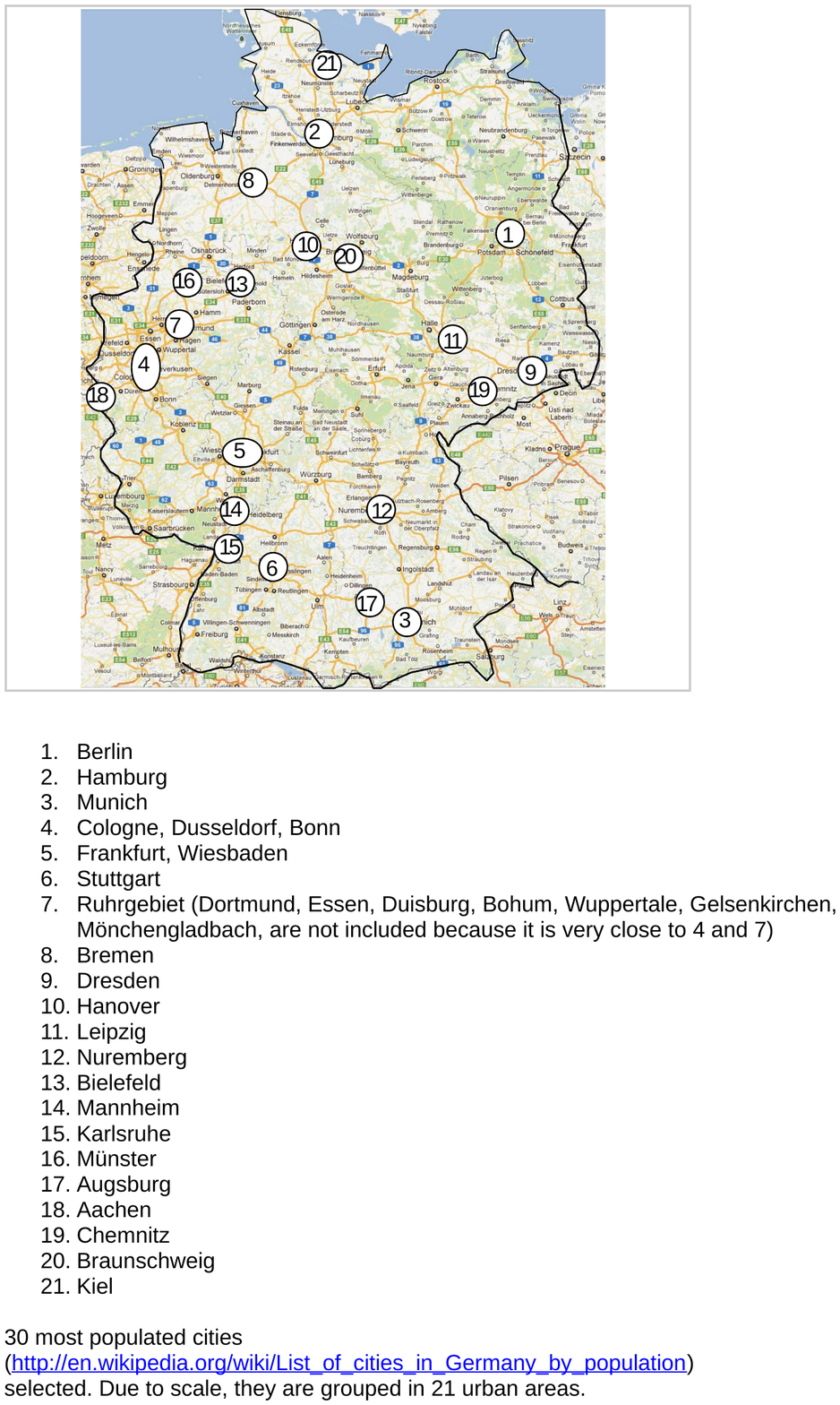}
\caption{A map of  Germany with major urban areas $\mathbf{U}$ shown by encircled numbers. }
\label{outlinemap}
\end{figure}

Plasmodium of \emph{ P. polycephalum} is cultivated in a plastic container, on paper kitchen towels moistened with 
still water, and fed with oat flakes. For experiments we use $120 \times 120$~mm polystyrene square Petri dishes
and 2\% agar gel (Select agar, by Sigma Aldrich) as a substrate. Agar plates, about 2-3~mm in depth, are cut in a shape of Germany.

We selected 21 most populated major urban areas listed below (see configuration of the areas in  
Fig.~\ref{outlinemap}a, which roughly corresponds to distribution of population densities by 
2009~\cite{statistics}):

\begin{multicols}{2}
\begin{enumerate}
\item Berlin
\item Hamburg
\item Munich
\item Cologne, including Dusseldorf, Bonn
\item Frankfurt, including Wiesbaden
\item Stuttgart
\item Dortmund area
\item Bremen
\item Dresden
\item Hanover
\item Leipzig
\item Nuremberg
\item Bielefeld
\item Mannheim
\item Karlsruhe
\item MŸnster
\item Augsburg
\item Aachen
\item Chemnitz
\item Braunschweig
\item Kiel
\end{enumerate}
\end{multicols}

Essen, Duisburg, Bohum, Wuppertale, Gelsenkirchen and M\"{o}nchengladbach are not included in the list because they is very close to either \Four and/or \Seven.

\begin{figure}[!tbp]
\centering
\includegraphics[width=0.8\textwidth]{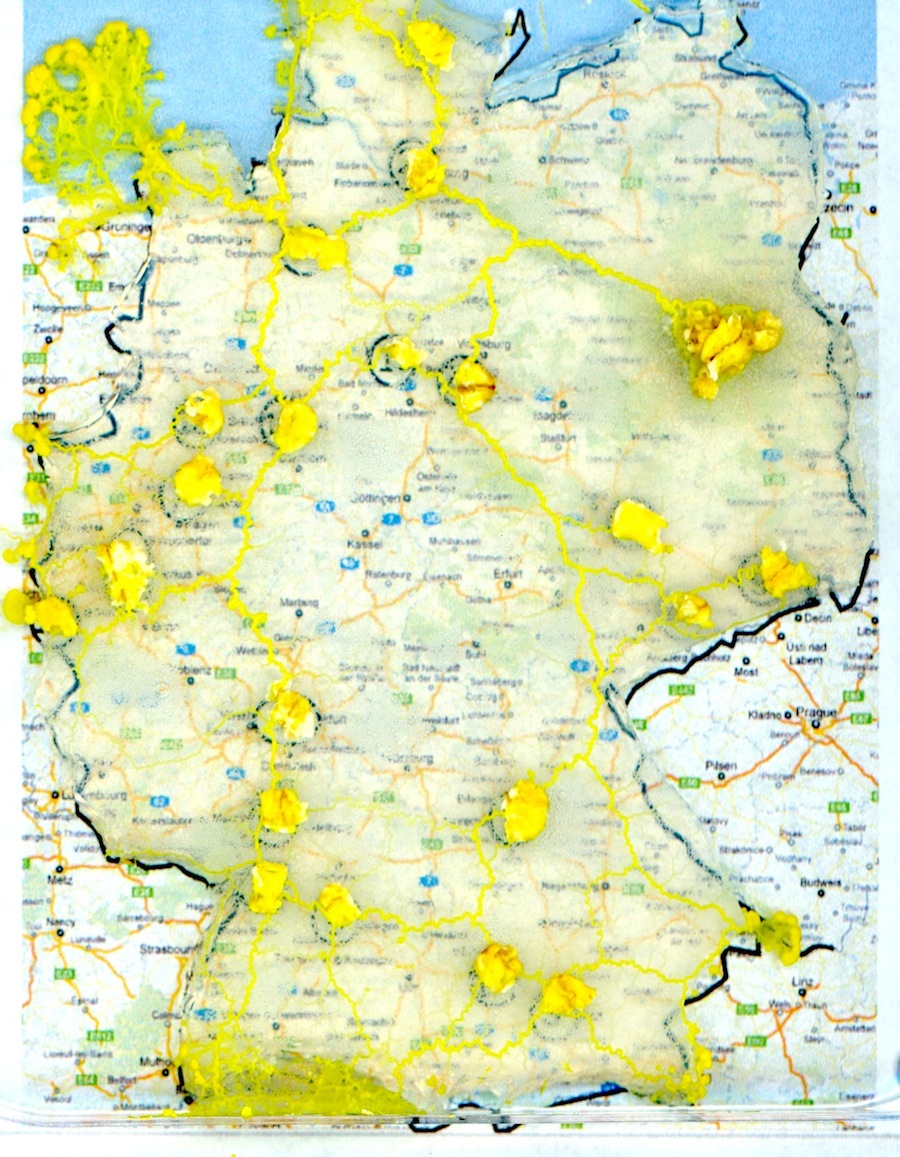}
\caption{A typical image of slime mould \emph{P. polycephalum} growing on a non-nutrient substrate and connecting oat flakes, which represent major urban areas $\mathbf{U}$ by a network of protoplasmic tubes.}
\label{schemPhys}
\end{figure}

To represent areas of $\mathbf{U}$ we place oat flakes  (each flake weights 9--13~mg and is 5--7~mm in diameter) in the positions of agar plate corresponding to the areas. At the beginning of each experiment an oat flake colonised by plasmodium (25--30~mg plasmodial weight) is placed in Berlin area.  We undertook 22 experiment. The Petri dishes with plasmodium are kept in darkness, at temperature 22-25$^\text{o}$C, except for observation and image recording. Periodically, the dishes are scanned with an Epson Perfection 4490 scanner and configurations of protoplasmic networks analysed. A typical image of experimental Petri dish with Germany-shaped gel plate colonised by 
\emph{P. polycephalum} is shown in Fig.~\ref{schemPhys}.

To generalise our experimental results we constructed a Physarum graph with weighted-edges. 
A Physarum graph  is a tuple ${\mathbf P} = \langle {\mathbf U}, {\mathbf E}, w  \rangle$, 
where $\mathbf U$ is a set of  urban areas, $\mathbf E$ is a set edges, and
$w: {\mathbf E} \rightarrow [0,1]$ associates each edge of $\mathbf{E}$ with  a probability (or weights).
For every two regions $a$ and $b$ from $\mathbf U$ there is an edge connecting $a$ and $b$ if a 
plasmodium's protoplasmic link is recorded at least in one of $k$ experiments, and the edge $(a,b)$ has a 
probability calculated as a ratio of experiments where protoplasmic link $(a,b)$ occurred in the total number 
of experiments $k=22$. For example,  if we observed a protoplasmic tube connecting areas $a$ and $b$ in 9 experiments, the weight of edge $(a,b)$ will be $w(a,b)=\frac{9}{30}$. We do not take into account the exact configuration of the protoplasmic tubes but merely their existence. 

Further we will be dealing with threshold Physarum graphs $\mathbf{P}(\theta)  = \langle  {\mathbf U}, T({\mathbf E}), w, \theta \rangle$. The threshold Physarum graph is obtained from Physarum graph 
by the transformation: $T({\mathbf E})=\{ e \in {\mathbf E}: w(e) > \theta \}$. That is all edges with weights less than or equal to  $\theta$ are removed.

\begin{figure}[!tbp]
\centering
\includegraphics[width=0.6\textwidth]{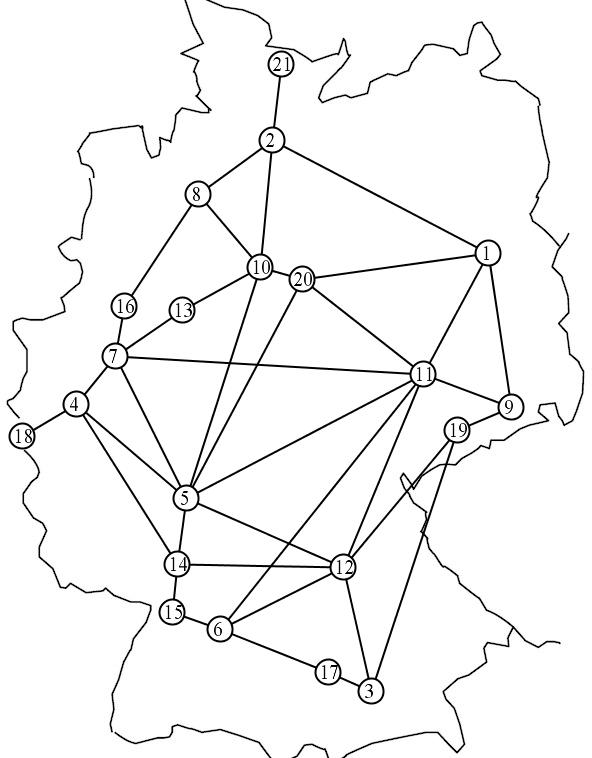}
\caption{Autobahn graph $\mathbf{H}$ of Germany.}
\label{motorway}
\end{figure}

To compare slime mould approximation of transport network in Germany with man-made autobahns we compare the generalised Physarum graph with the autobahn graph $\mathbf H$. The autobahn graph is derived as follows. Let $\mathbf U$ be a set of urban regions/cities; for any two regions $a$ and $b$ from $\mathbf U$, the nodes $a$ and $b$ are connected by an edge $(ab)$ if there is an autobahn starting in vicinity of $a$, passing in vicinity of $b$, and not passing in vicinity of any other urban area $c \in \mathbf U$. In the case of branching --- that is, a autobahn starts in 
$a$, goes in the direction of $b$ and $c$, and at some point branches towards $b$ and $c$ --- we then add two separate edges $(ab)$ and $(ac)$ to the graph $\mathbf H$.  The autobahn graph is planar (Fig.~\ref{motorway}).

We also analyse autobahn and Physarum graphs in a context of planar proximity graphs. 
A planar graph consists of nodes which are points of the  Euclidean plane and edges which are straight segments connecting the points. A planar proximity graph is a planar graph where two points are connected by an edge if they are close in some sense, and no edges intersect. A pair of points is assigned a certain neighbourhood, and points of the pair are connected by an edge if their neighbourhood is empty.  Here we consider the most common proximity graph as follows.
\begin{itemize}
\item $\mathbf{GG}$: Points $a$ and $b$ are connected by an edge in the Gabriel Graph $\mathbf{GG}$ if
disc with diameter $dist(a,b)$ centred in middle of the segment $ab$ is 
empty~\cite{gabriel_sokal_1969,matula_sokal_1984}.
\item $\mathbf{RNG}$: Points $a$ and $b$ are connected by an edge in the Relative Neighbourhood Graph $\mathbf{RNG}$ if no other point $c$ is closer
to $a$ and $b$ than $dist(a,b)$~\cite{toussaint_1980}.
\item $\mathbf{MST}$: The Euclidean minimum spanning tree (MST)~\cite{nesetril} is a connected acyclic graph which has minimum possible sum of edges' lengths. 
\end{itemize}
In general, the graphs relate as
$\mathbf{MST} \subseteq \mathbf{RNG}  \subseteq\mathbf{GG}$~\cite{toussaint_1980,matula_sokal_1984,jaromczyk_toussaint_1992}; this is called Toussaint hierarchy.

\section{Strategies of colonisation}
\label{colonisation}

Being inoculated in \One slime mould of \emph{P. polycephalum}  develops by one of two scenarios: anti-clockwise propagation along boundaries of Germany, in the directions west -- south -- east -- north; and, clockwise propagation, in the directions south -- east -- north -- west. In 13 of 22 experiments plasmodium propagates anti-clockwise, in 9 experiments --- clockwise.

\begin{figure}[!tbp]
\centering
\subfigure[48~hr]{\includegraphics[width=0.44\textwidth]{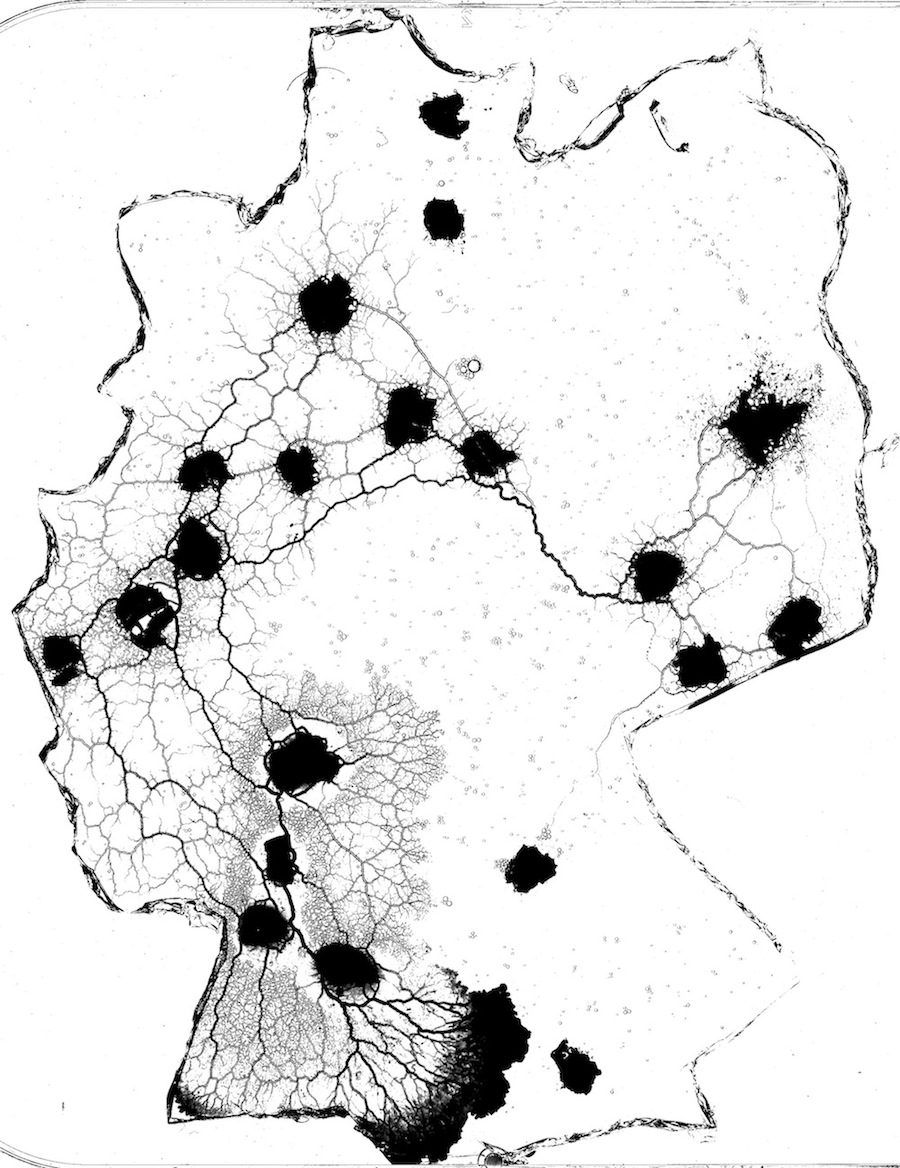}}
\subfigure[72~hr]{\includegraphics[width=0.44\textwidth]{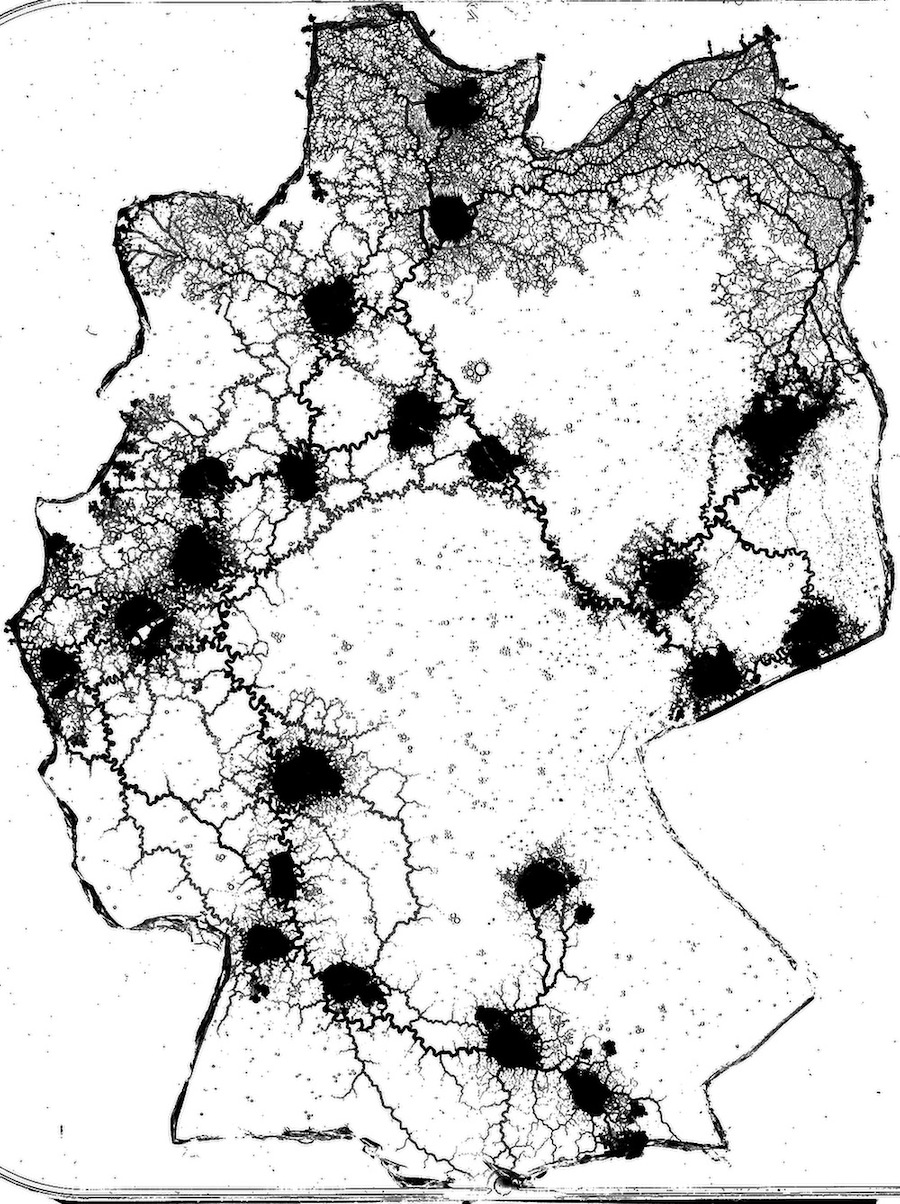}}
\caption{Illustration of anti-clockwise propagation. Experimental laboratory snapshots of \emph{P. polycephalum} colonising urban areas of $\mathbf{U}$. The snapshots are taken (a)~48~h and (b)~72~h after inoculating 
the plasmodium in \One. }
\label{exm08}
\end{figure}

Anti-clockwise scenario of colonisation is shown in 
Fig.~\ref{exm08}. The plasmodium is inoculated in Berlin. It propagates to 
and colonises \OneOne and \Nine simultaneously. Then it builds protoplasmic tubes 
connecting \Nine and \OneOne with  \OneNine (Fig.~\ref{exm08}a). At the same
time, the plasmodium grows from \OneOne to   \TwoZero, and then occupies almost 
all urban areas from \Eight in the north to \OneSeven in the 
south (Fig.~\ref{exm08}a). By the 3rd day after inoculation the plasmodium propagates
from \OneSeven to   \Three and \OneTwo, and from \Eight  to  \Two and  from  \Two  to 
\TwoOne   (Fig.~\ref{exm08}b).

\begin{figure}[!tbp]
\centering
\subfigure[48~hr]{\includegraphics[width=0.44\textwidth]{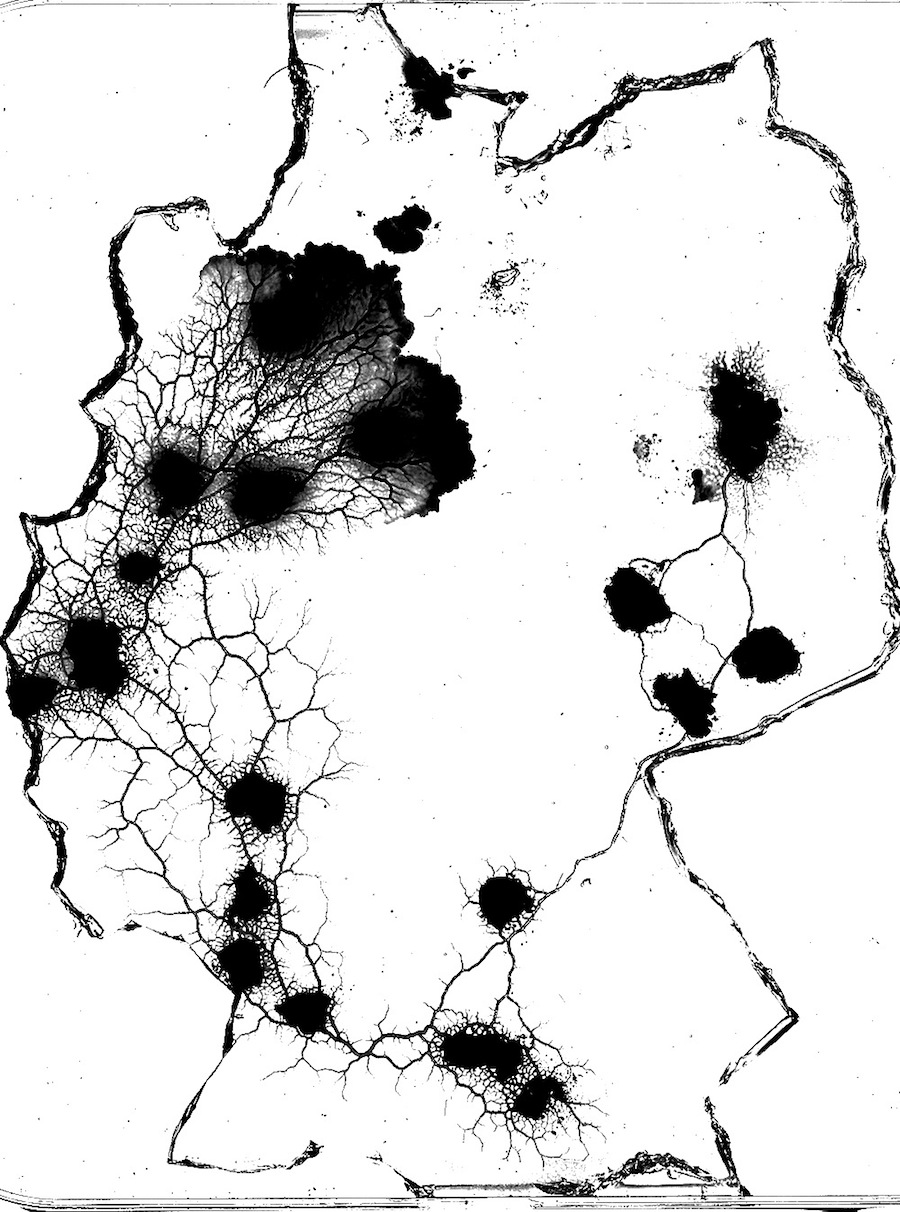}}
\subfigure[72~hr]{\includegraphics[width=0.44\textwidth]{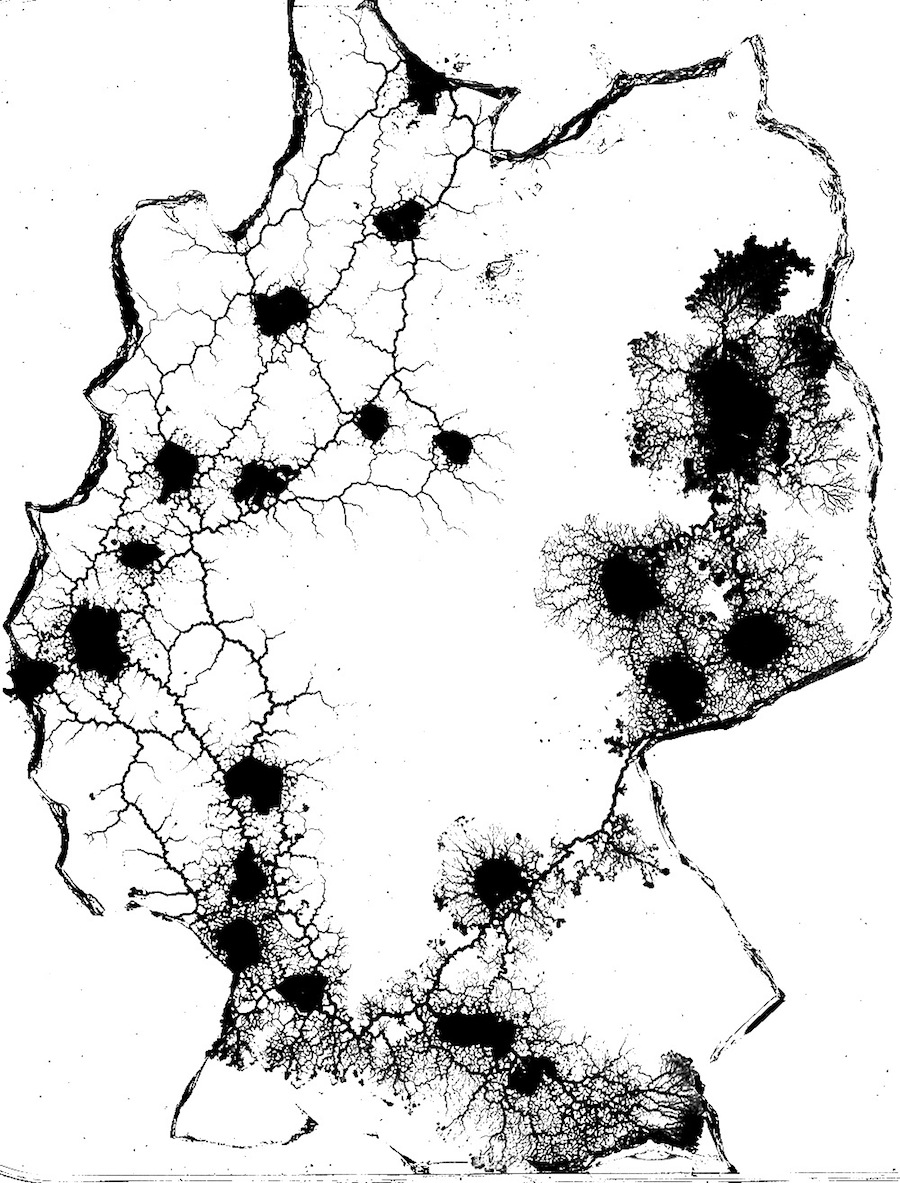}}
\caption{Illustration of clockwise propagation. Experimental laboratory snapshots of \emph{P. polycephalum} colonising urban areas of $\mathbf{U}$. The snapshots are taken (a)~48~h and (b)~72~h after inoculating 
the plasmodium in \One.}
\label{exm08}
\end{figure}

Clockwise (south-east-north) scenario is illustrated in Fig.~\ref{exm08}. The slime mould 
propagates from \One to \OneOne and \Nine simultaneously. Then it colonises \OneNine, 
forming protoplasmic tubes (\OneOne -- \OneNine) and (\Nine -- \OneNine). It moves 
further south and occupies the oat flake corresponding to urban area \OneTwo 
(Fig.~\ref{exm08}a). In its subsequent development, by 48~hr after
inoculation in \One, the plasmodium
develops a chain of protoplasmic tubes linking
\OneTwo, \OneSeven, \Three in the south, \Six, \OneFive, \OneFour and 
\Five in the south-west, and \OneEight, \Four,  \Seven, \OneSix,
\OneThree in the west (Fig.~\ref{exm08}a).
The colonisation of urban areas $\mathbf{U}$ is completed by 
the slime mould by 72~h after inoculation, when it develops protoplasmic 
tubes connecting \Eight and \OneZero with \Two and \TwoOne (Fig.~\ref{exm08}b).

\begin{figure}[!tbp]
\centering
\subfigure[48~hr]{\includegraphics[width=0.44\textwidth]{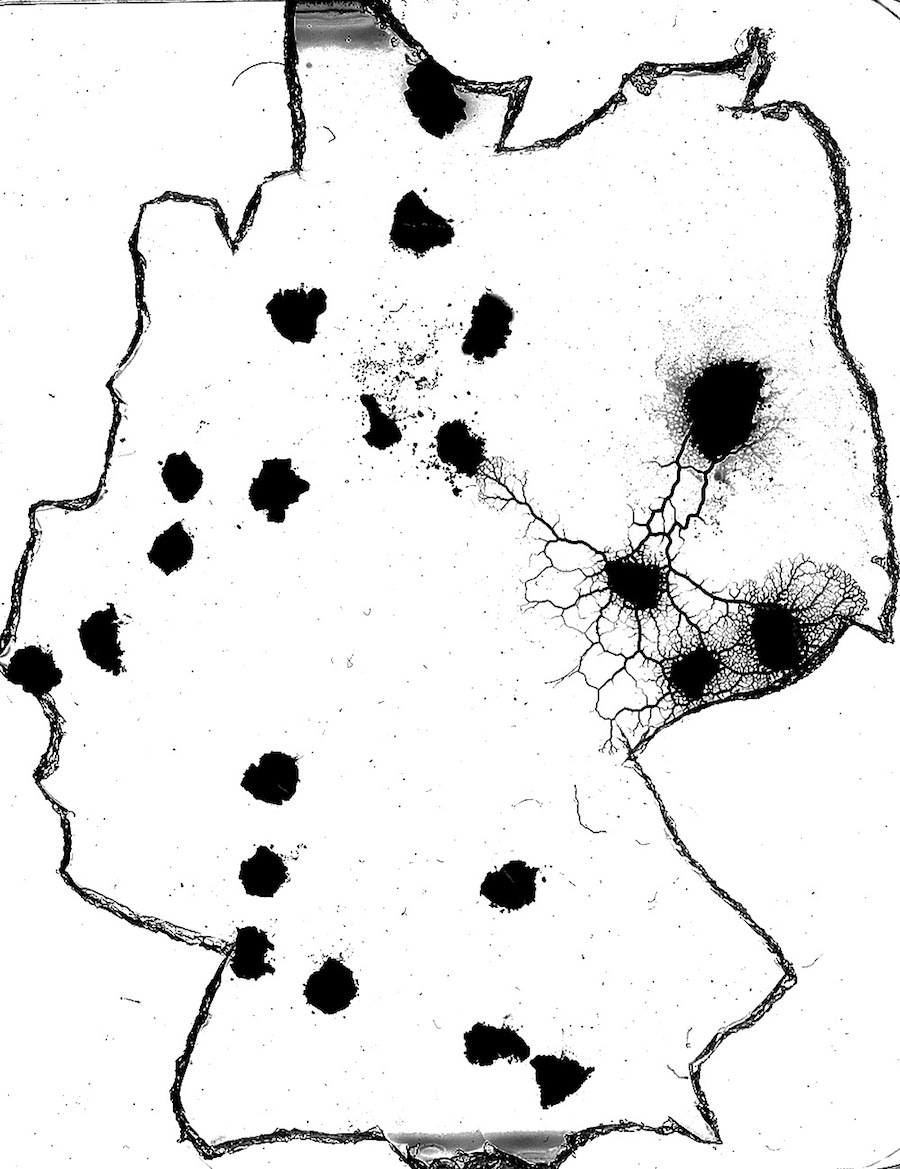}}
\subfigure[72~hr]{\includegraphics[width=0.44\textwidth]{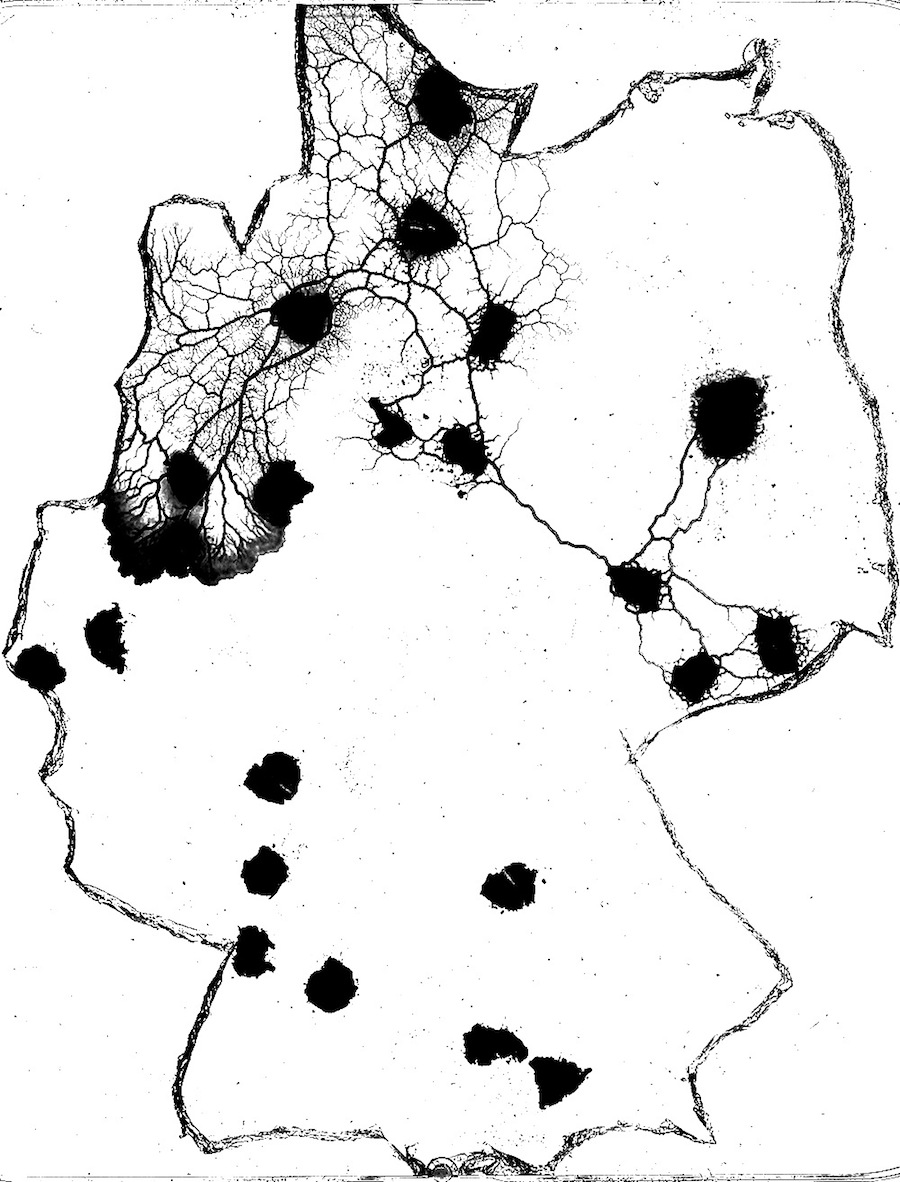}}
\subfigure[96~hr]{\includegraphics[width=0.44\textwidth]{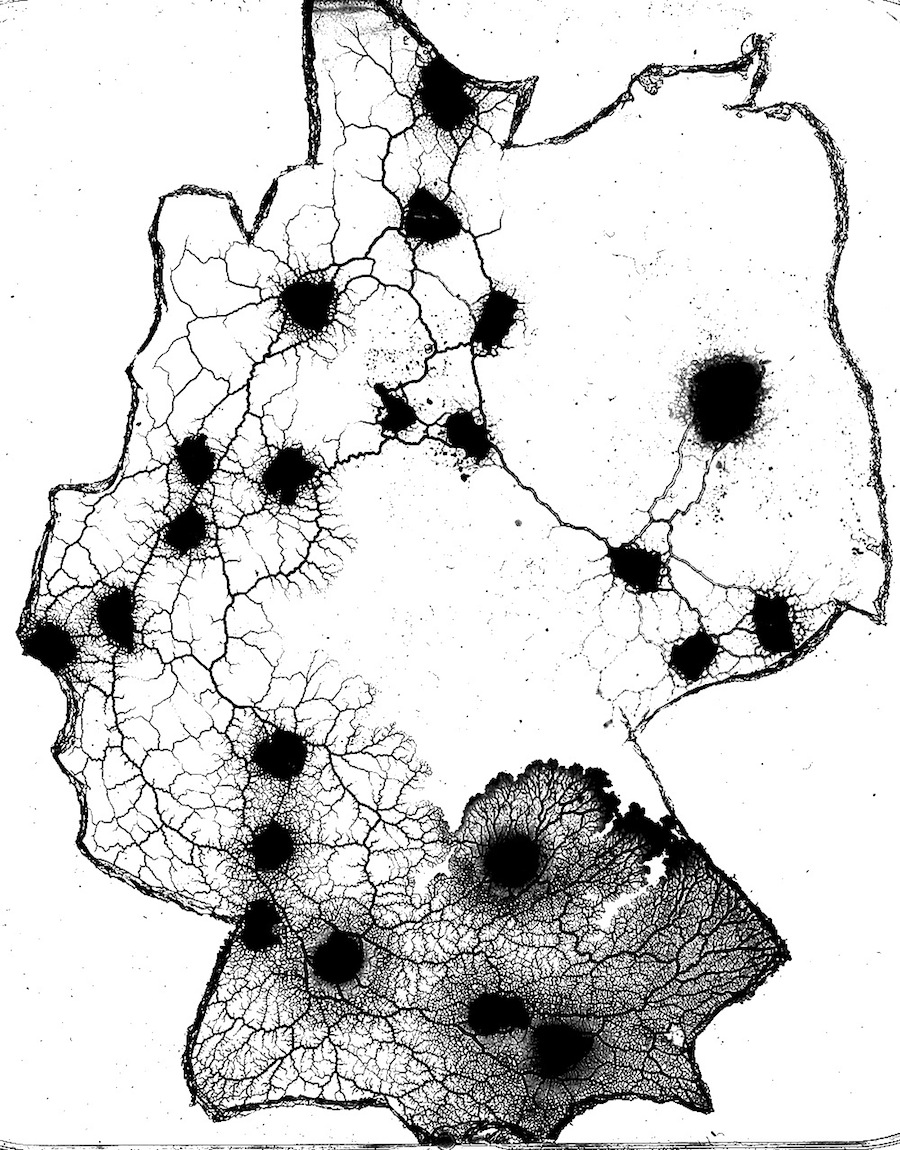}}
\subfigure[Scheme]{\includegraphics[width=0.44\textwidth]{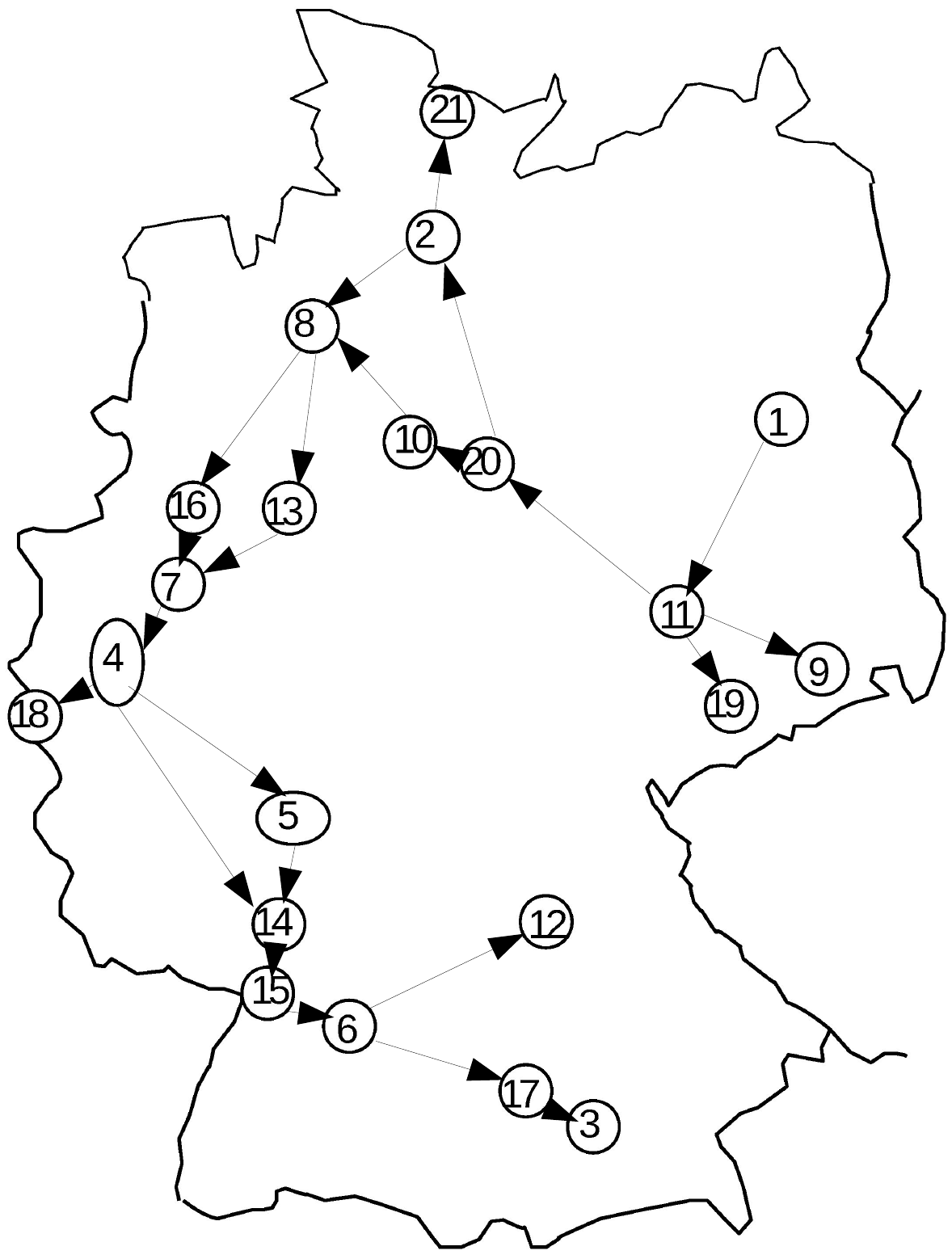}}
\caption{Illustration of decision-making by the plasmodium. The snapshots are taken (a)~48~h and (b)~72~h after inoculating  the plasmodium in \One.}
\label{exm12}
\end{figure}

A basic decision-making process implemented by \emph{P. polycephalum}
is shown in Fig.~\ref{exm12}.  In the first two days after being inoculated in \One 
the plasmodium propagates to \OneOne. It then branches from \OneOne to 
\TwoZero, \OneNine and \Nine simultaneously (Fig.~\ref{exm12}a).  At this moment the plasmodium
has three options:
\begin{enumerate}
\item propagate from \OneNine to \OneTwo, \OneSeven and \Three 
and then westward and towards north,
\item propagate from \TwoZero westward and then southward,
\item implement options 1 and 2 in pallel.
\end{enumerate}
In this particular experimental-laboratory example the plasmodium chooses option 2, 
because the concentration of nutrients (detected via chemo-attractants) is higher in the region 
of \TwoZero than around \OneNine. See scheme of propagation in Fig.~\ref{exm12}d.
It propagates from \TwoZero to \OneZero and \Two; from \Two to \TwoOne;
from \Two to \Eight;  from \Eight to \OneSix and \OneThree; 
and, from \OneSix and \OneThree to \Seven (Fig.~\ref{exm12}b).
On the fourth day after inoculation the plasmodium colonises \Four and \OneEight;
builds protoplasmic veins linking \Four to \Five and \OneFour; \OneFour to 
\OneFive; \OneFive to \Six. Snapshot Fig.~\ref{exm12}c shows 'moment' when 
the plasmodium just colonised \Three, \OneTwo and \OneSeven and starts to 
develop growing zones to explore the space around newly occupied areas. 
The plasmodium also is detecting shortest ways, and  evaluating feasibility of 
propagation, towards \OneNine.

\section{Physarum graphs}
\label{physarumgraphs}

\begin{figure}[!tbp]
\centering
\subfigure[$\theta=\frac{1}{22}$]{\includegraphics[width=0.32\textwidth]{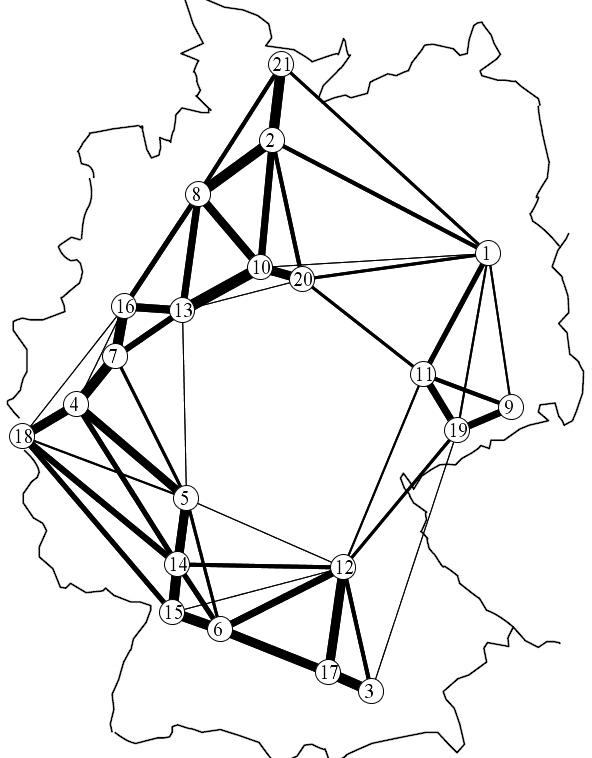}}
\subfigure[$\theta=\frac{8}{22}$]{\includegraphics[width=0.32\textwidth]{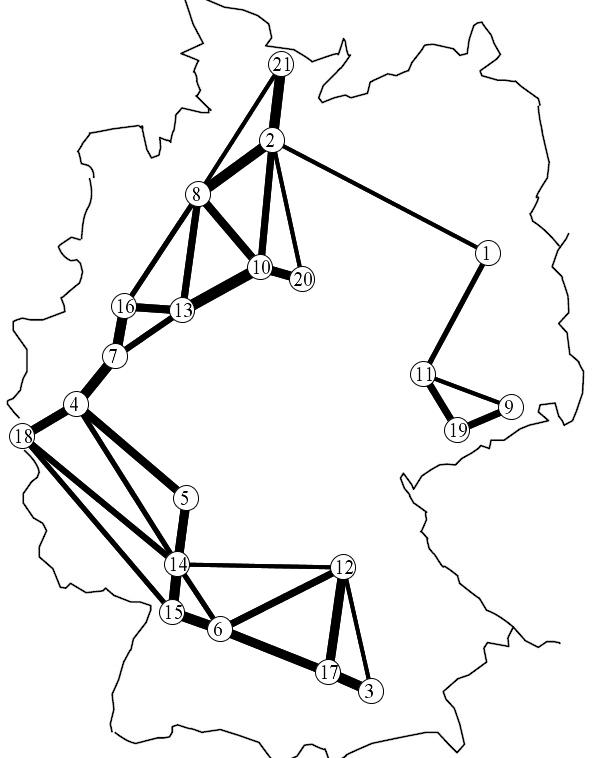}}
\subfigure[$\theta=\frac{9}{22}$]{\includegraphics[width=0.32\textwidth]{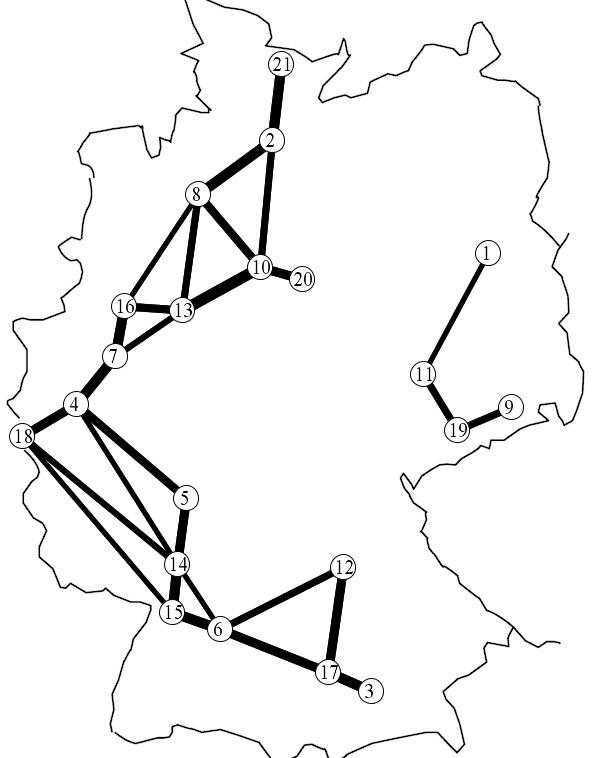}}
\subfigure[$\theta=\frac{11}{22}$]{\includegraphics[width=0.32\textwidth]{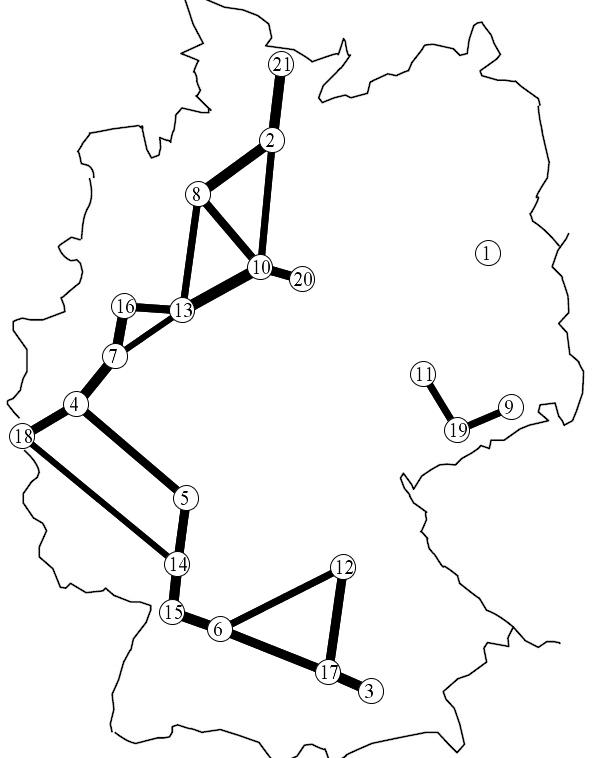}}
\subfigure[$\theta=\frac{14}{22}$]{\includegraphics[width=0.32\textwidth]{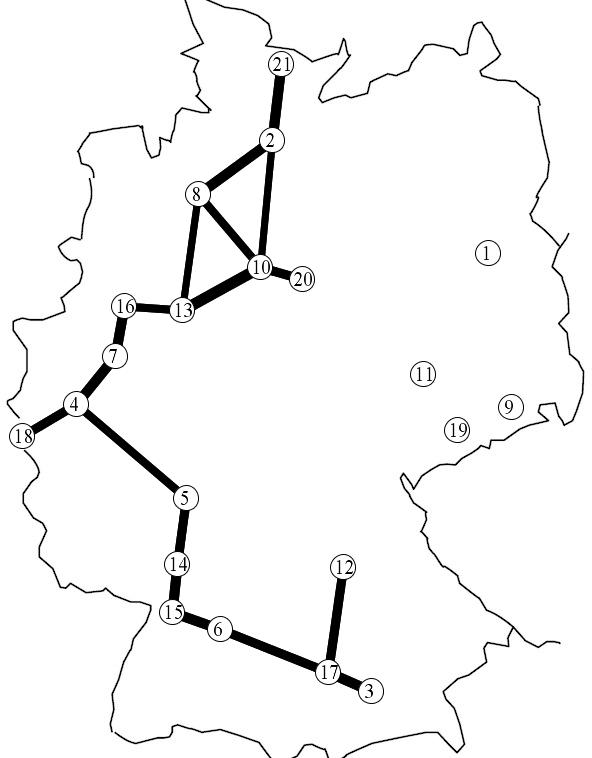}}
\subfigure[$\theta=\frac{15}{22}$]{\includegraphics[width=0.32\textwidth]{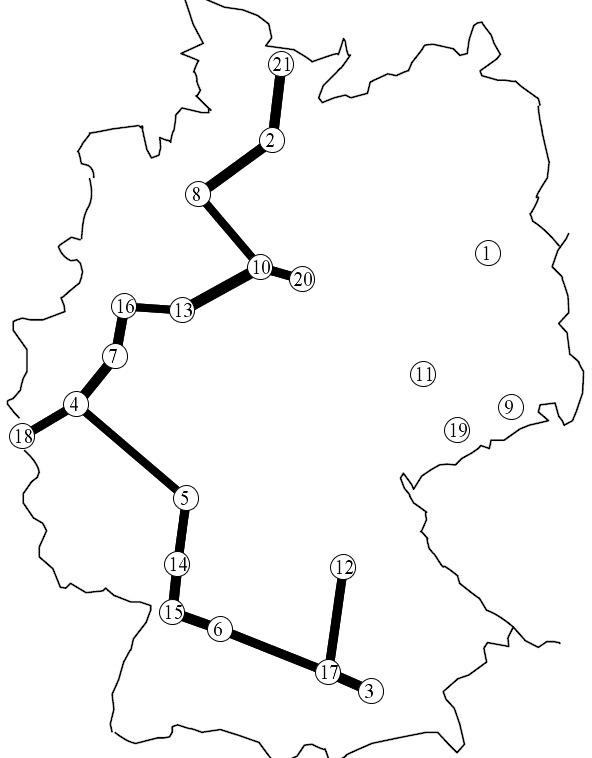}}
\subfigure[$\theta=\frac{20}{22}$]{\includegraphics[width=0.32\textwidth]{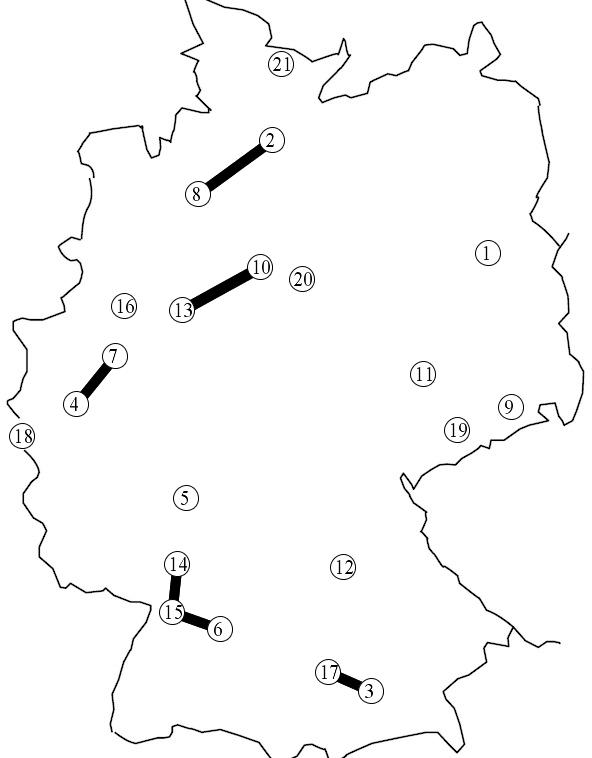}}
\subfigure[$\theta=\frac{21}{22}$]{\includegraphics[width=0.32\textwidth]{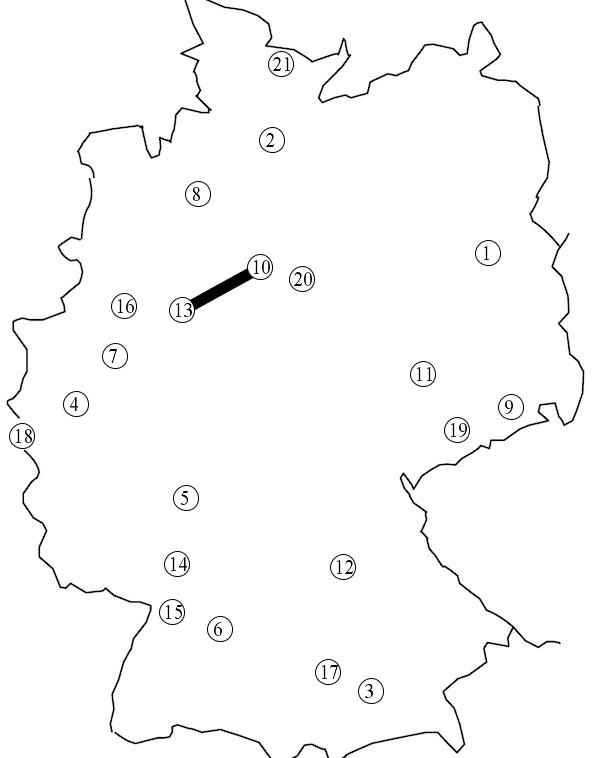}}
\caption{Generalized Physarum graphs $\mathbf{P}(\theta)$ for selected values of $\theta$.}
\label{physarumgraphs}
\end{figure}

Examples of threshold Physarum graphs for some values of $\theta$, are shown in Fig.~\ref{physarumgraphs}. 
The raw Physarum graph $\mathbf{P}(\frac{1}{22})$ is non-planar (Fig.~\ref{physarumgraphs}). With increase of 
$\theta$ -- the higher is $\theta$ of  $\mathbf{P}(\theta)$ the more often edges of $\mathbf{P}(\theta)$  appear in laboratory experiments --- the threshold Physarum graphs undergo the following transformations: 
\begin{itemize}
\item $\frac{1}{22} \leq \theta \leq \frac{4}{22}$: graph $\mathbf{P}(\theta)$ is non-planar and connected (Fig.~\ref{physarumgraphs}a).
\item $\frac{5}{22} \leq \theta \leq \frac{8}{22}$: graph $\mathbf{P}(\theta)$ is planar and connected (Fig.~\ref{physarumgraphs}b).s  
\item $\theta=\frac{9}{22}$: graph $\mathbf{P}(\theta)$ splits into two disconnected components, one consists of 
urban areas \One, \Nine, \OneOne and \OneNine, another include the remaining areas (Fig.~\ref{physarumgraphs}c),
\item $\theta=\frac{11}{22}$: \One becomes isolated vertex (Fig.~\ref{physarumgraphs}d).
\item  $\theta=\frac{14}{22}$:  \One, \Nine, \OneOne and \OneNine become isolated vertices  
(Fig.~\ref{physarumgraphs}e).
\item $\theta=\frac{15}{22}$:  graph $\mathbf{P}(\theta)$  becomes acyclic  (Fig.~\ref{physarumgraphs}f). 
\end{itemize}  

\begin{proposition}
Slime mould P. polycephalum imitates 1947 year separation of Germany onto East Germany 
and West Germany.
\end{proposition}

See Fig.~\ref{physarumgraphs}c. When we consider only edges represented by protoplasmic tubes in over 
41\% of laboratory experiments Physarum graph becomes split into two disconnected components. The eastern component is a chain of urban areas  \One -- \OneOne -- \OneNine -- \Nine lies exactly in the territory of former Eastern 
Germany (Fig.~\ref{physarumgraphs}c). The construction of German transport network was interrupted by the Second World War and resumed only in 1953.  Over half of existing autobahns in the West German network had been constructed by 1975~\cite{rothengatter_2005} well before fall of the Berlin Wall. The 'weak' fragments, i.e. those which are removed from Physarum graphs with increase of $\theta$, of motorway network between West and East Germany are the ones which link now territories of East and West Germany. Another contributing factor to the weak links exposed in laboratory experiments with \emph{P. polycephalum} could be that car ownership and roads were symbols of freedom and democracy in West Germany, while East Germany characterised with low level of car ownership and poorly maintained roads.

Further increase of $\theta$ to $\frac{16}{22}$
leads to separation of the graph into the following 
disconnected components: isolated vertices  \One, \Nine, \OneOne and \OneNine, 
two three link chains \TwoOne -- \Two -- \Eight  and 
\TwoZero -- \OneZero -- \OneThree, one four link chain 
\OneSix -- \Seven -- \Four -- \OneEight and the five link chain
\Five -- \OneFour -- \OneFive -- \Six -- \OneSeven -- \Three and the branch
\OneSeven -- \OneTwo. 

\begin{finding}
Transport links presented in over 68\% of laboratory experiments form a connected component 
consisting of  a chain \TwoOne -- \Two -- \Eight -- \OneZero --
\OneThree -- \OneSix -- \Seven -- \Four -- \Five -- \OneFour -- \OneFive -- \Six -- \OneSeven --
\Three with three branches: \OneZero -- \TwoZero
, \Four -- \OneEight, \OneSeven -- \OneTwo.
\end{finding}

See Fig.~\ref{physarumgraphs}f. The slime mould representations reflects thus an ever existing imbalance. 
Historically the West of Germany has always been richer than the East, the highest concentration of 
steel and coal-mining industry and engineering was situated in the Ruhrgebiet (Essen, Dortmund etc.).

\section{Autobahns versus protoplasmic networks}
\label{comparison}

\begin{finding}
The only autobahn links presented by P. polycephalum in over 90\% of laboratory trials are 
(\Two -- \Eight), (\OneZero -- \OneThree), (\Four -- \Seven), (\OneFour -- \OneFive -- \OneSix), and  (\OneSeven -- \Three). 
\end{finding}

All these links are parts of the so-called Reichsautobahn built by 1940 during Hitler Germany~\cite{autobahns_1940}. The only transport link connecting \OneZero to \OneThree is represented by protoplasmic tubes in 95\% of all laboratory experiments (Fig.~\ref{physarumgraphs}jh).

\begin{figure}[!tbp]
\centering
\subfigure[$\mathbf{H} \cap \mathbf{P}(\frac{1}{22})$]{\includegraphics[width=0.32\textwidth]
{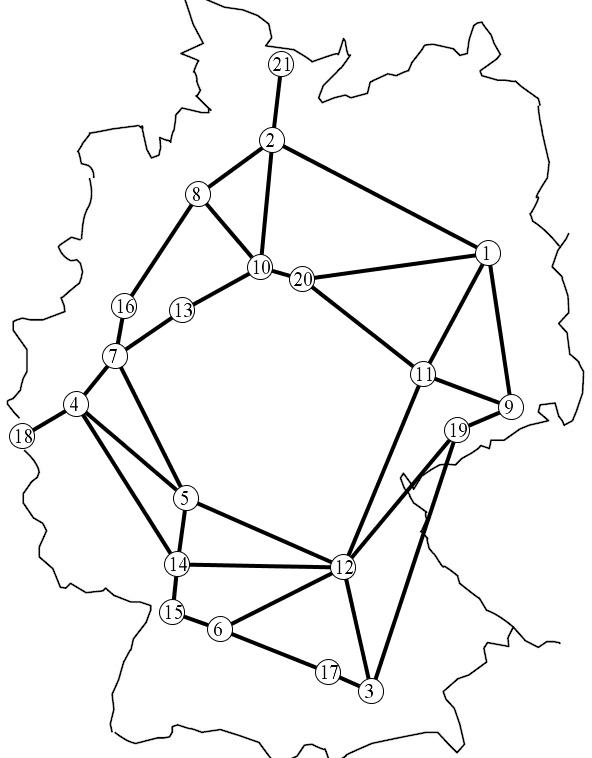}}
\subfigure[$\mathbf{H} \cap \mathbf{P}(\frac{7}{22})$]{\includegraphics[width=0.32\textwidth]
{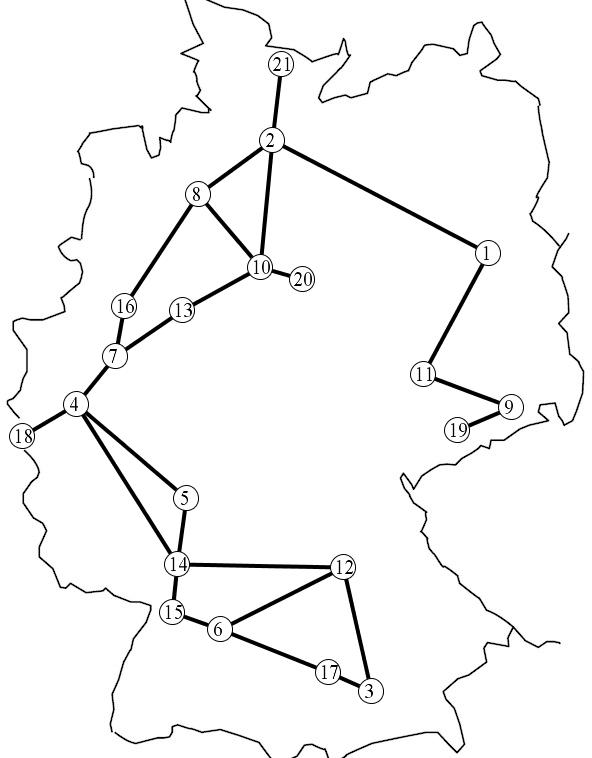}}
\subfigure[$\mathbf{H} \cap \mathbf{P}(\frac{15}{22})$]{\includegraphics[width=0.32\textwidth]
{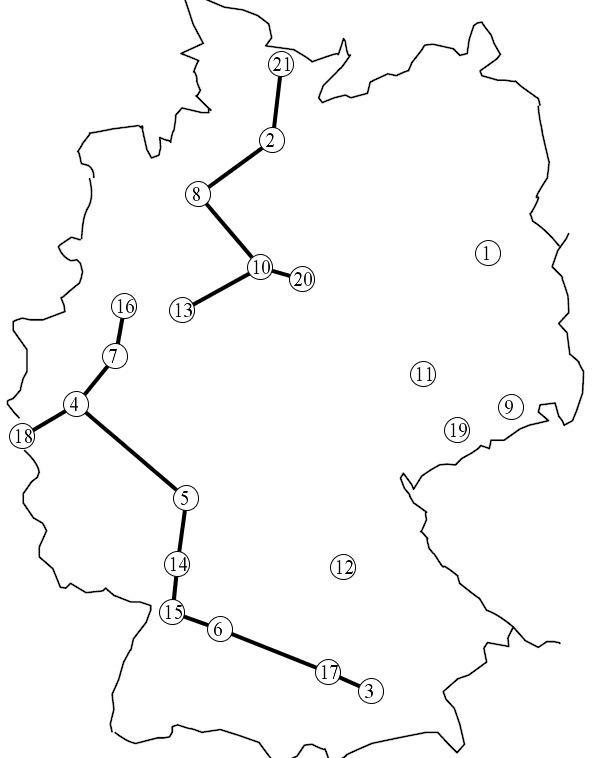}}
\caption{Intersection of  autobahn graph $\mathbf{H}$ with Physarum graphs 
(a)~$\mathbf{P}(\frac{1}{22})$,
(b)~$\mathbf{P}(\frac{7}{22})$, 
(c)~$\mathbf{P}(\frac{15}{22})$.
}
\label{intersetmwayphysarum}
\end{figure}

\begin{finding}
Autobahn links (\Five -- \TwoZero),   (\Five -- \OneZero), (\Five -- \OneOne) and  (\Six -- \OneOne) are never represented by protoplasmic tubes of P. polycephalum.
\end{finding}

As we can see in Fig.~\ref{intersetmwayphysarum}a 'raw' Physarum graph, consisting of edges, which are represented by protoplasmic tubes in at least two experiments, is almost a sub-graph of the autobahn graph apart of edges 
(\Five -- \TwoZero),   (\Five -- \OneZero), (\Five -- \OneOne) and  (\Six -- \OneOne). The intersection of Physarum graph, which edges appear in at least eight of 22 experiments, and the autobahn graph remains a connected graph (Fig.~\ref{intersetmwayphysarum}b). 
The intersection  $\mathbf{H} \cap \mathbf{P}(\frac{15}{22})$ consists of 
\begin{itemize}
\item four isolated vertices: 
\One, \Nine, \OneOne, \OneTwo and \OneNine;
\item chain \TwoOne -- \Two -- \Eight -- \OneZero -- \OneThree with branch \OneZero -- \TwoZero; 
\item chain \OneSix -- \Seven -- \Four -- \Five -- \OneFour -- \OneFive -- \Six -- \OneSeven -- \Three with branch \Four -- \OneEight (Fig.~\ref{intersetmwayphysarum}c).
\end{itemize}

\begin{table}
\caption{Comparison of autobahn graph $\mathbf{H}$ and Physarum graph $\mathbf{P}(\frac{8}{22})$ using 
standard measures and indices. Rows are in ascending order of absolute value of mismatch.}
\begin{tabular}{p{8cm}||l|l|p{2cm}}
Measure  $\mu$		& $\mathbf{H}$ 	& $\mathbf{P}(\frac{7}{22})$ & Mistmatch, $1- \frac{\mu(\mathbf{H})}{\mu(\mathbf{P}(\frac{7}{22}))}$ \\ \hline \hline
				& 		  	 &			&	\\ 	
Randi\'{c} index~\cite{randic_1975}: $\sum_{ij} C_{ij}*(\frac{1}{\sqrt(d_i*d_j)})$, where $C_{ij}$ is a connectivity matrix 
& 19.88 & 20.06  &   -0.01  \\ \hline
Average link length	         & 0.44	   	& 0.45  &   0.02  \\ \hline
Average degree                     & 3.61	            	& 3.24  &   -0.11  \\ \hline
Connectivity: number of edges divided by number of nodes  			&  1.81	  	& 1.62   & -0.12   \\ \hline
$\Pi$-index: The relationship between the total length of the graph $L(G)$ and the distance along its diameter $D(d)$
\cite{Ducruet_2012}, $\Pi=\frac{L(G)}{D(d)}$	          & 85.59  		& 	74.21  &    -0.15 \\ \hline
Average cohesion: let $\overline{d}$ be an average degree of a graph $\mathbf G$ and $\nu_{ij}$ is a number of common neighbours of nodes $i$ and $j$, and $d_i$ is a degree of node $i$, then cohesion  $\kappa_{ij}$ between nodes $i$ and $j$ is calculated as $\kappa_{ij}=\frac{\nu_{ij}}{d_i + d_j}$
 		         & 0.19           	& 0.23  &  0.17   \\  \hline
Average shortest path between any two nodes, in nodes		& 2.49		   	& 4.03  &   0.38  \\ \hline
Harary index~\cite{plavsic_1993}: $\frac{1}{2} \sum_{ij} \xi(D)_{ij}$, where
where $i$ and $j$ are indices of a graph nodes, $D$ is a graph distance matrix, where $D_{ij}$ is a length of a shortest path between $i$ and $j$, $\xi(D)_{ij}= D^{-1}_{ij}$ if $i \neq j$ and 0, otherwise.		 & 293.50 		& 199.85  &   -0.47  \\ \hline
Average shortest path between any two nodes, in normalised lengths                & 0.98	  	& 1.89  &  0.48   \\ \hline
Diameter (longest shortest path between any two nodes), in nodes	          & 5		   	& 10  &  0.5   \\  \hline
Diameter (longest shortest path between any two nodes), in normalised lengths	          & 2.26	 	& 5.10  &    0.55 \\ \hline
\end{tabular}
\label{comparisontable}
\end{table}

Let us compare Physarum graph $\mathbf{P}(\frac{7}{22})$ with the autobahn graph based on several integral measures listed in Tab.~\ref{comparisontable}.  For each measure $\mu$ we calculate a mismatch between the Physarum and autobahn graphs as $1- \frac{\mu(\mathbf{H})}{\mu(\mathbf{P}(\frac{7}{22}))}$.  The graphs show 
very good match in 
\begin{itemize}
\item  Randi\'{c} index: the index is calculated as $\sum_{ij} C_{ij}*(\frac{1}{\sqrt(d_i*d_j)})$, where $C_{ij}$ is a connectivity matrix~\cite{randic_1975};
\item average link length;
\item average degree;
\item  connectivity:  number of edges divided by number of nodes,
\item $\Pi$-index;
\item average cohesion, which somehow reflects a neighbourhood wholeness,
\end{itemize}

\begin{proposition}
Slime mould P. polycephalum imitates well the autobahn network in 
terms of reachability, average travel time between geographically close urban areas 
and a fault-tolerance. 
\end{proposition}

The reachability is  expressed as connectivity (Tab.~\ref{comparisontable}). The average travel time between 
geographical close urban areas could be measured via average link length. The number of alternative 
routes is measured via graphs' branching properties reflected in the   Randi\'{c} index~\cite{liu_2005}. 
The slime mould does not represent
well the following properties of German transport networks:   
diameter (longest path between any two vertices, measured in nodes or normalised length of edges),  Harary index,
and average shortest path between a pair of vertices (Tab.~\ref{comparisontable}).

\section{Slime mould, autobahns and proximity graphs}
\label{proximity}

\begin{figure}[!tbp]
\centering
\subfigure[${\mathbf{GG}}$]{\includegraphics[width=0.32\textwidth]{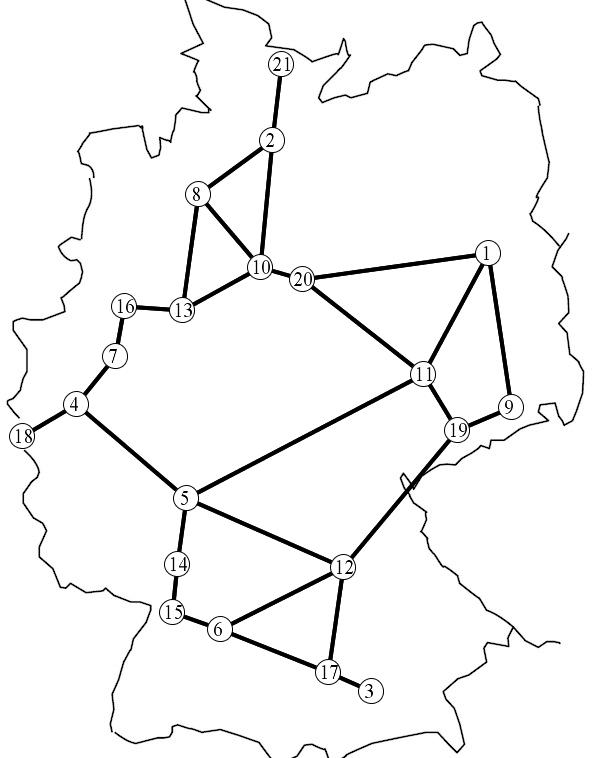}}
\subfigure[${\mathbf{RNG}}$]{\includegraphics[width=0.32\textwidth]{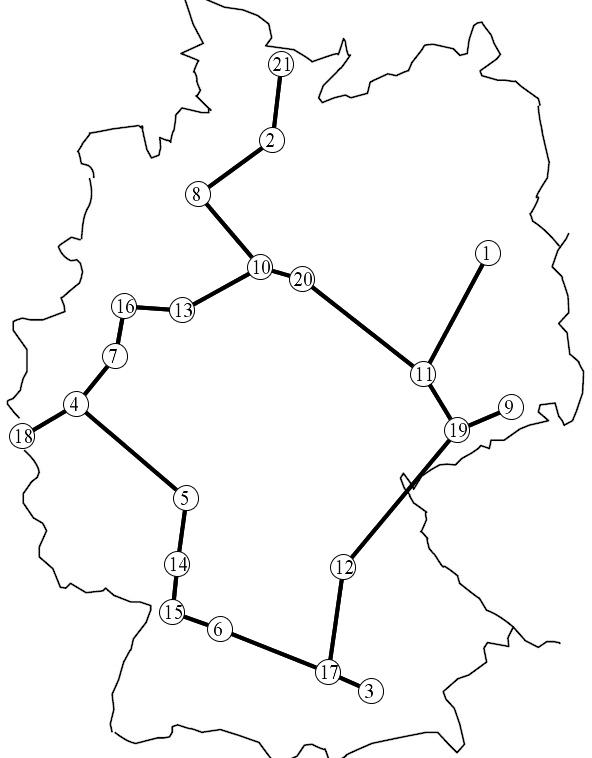}}
\caption{Proximity graphs constructed on sites of ${\mathbf U}$. 
(a)~Gabriel graph. (b)~Relative neighbourhood graph.}
\label{proximity}
\end{figure}

\begin{figure}[!tbp]
\centering
\subfigure[$\mathbf{ST}_1$]{\includegraphics[width=0.2\textwidth]{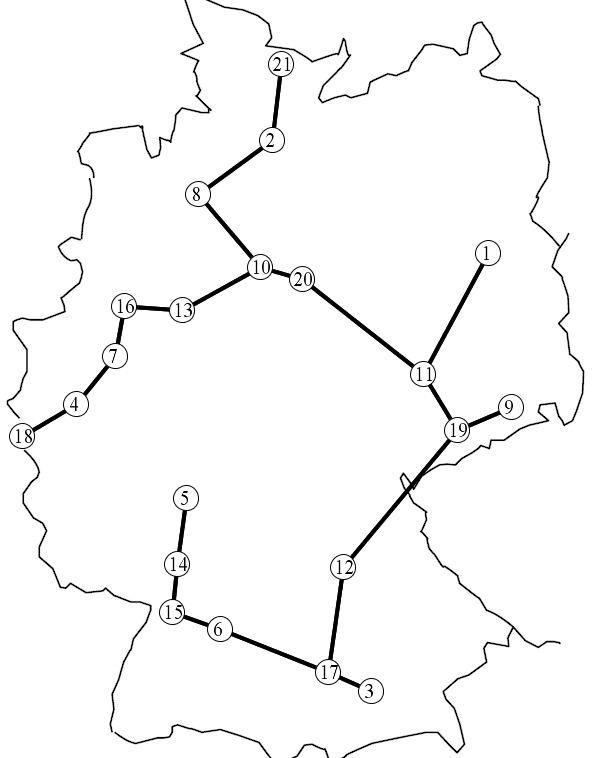}}
\subfigure[$\mathbf{ST}_2$]{\includegraphics[width=0.2\textwidth]{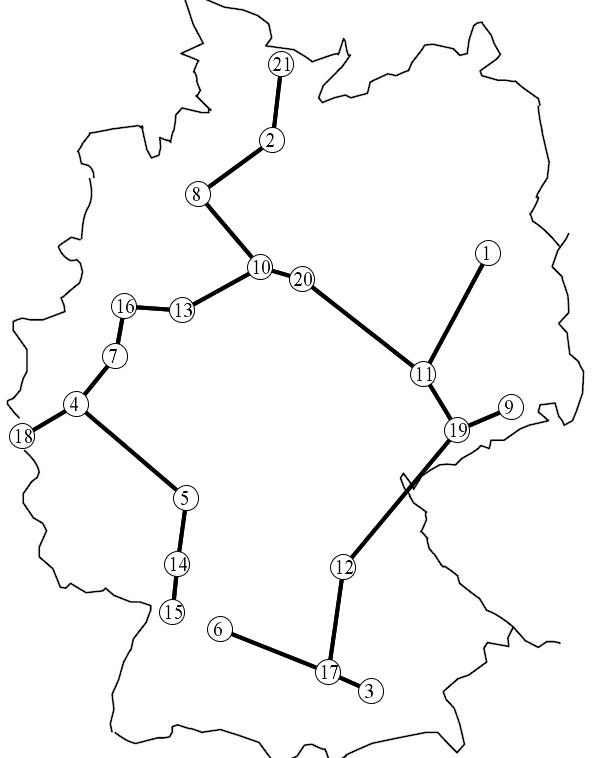}}
\subfigure[$\mathbf{ST}_3$]{\includegraphics[width=0.2\textwidth]{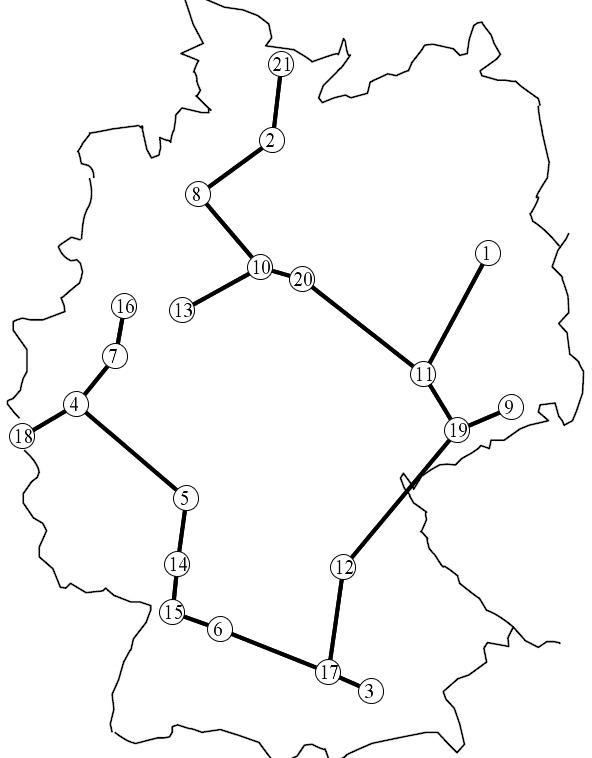}}
\subfigure[$\mathbf{ST}_4$]{\includegraphics[width=0.2\textwidth]{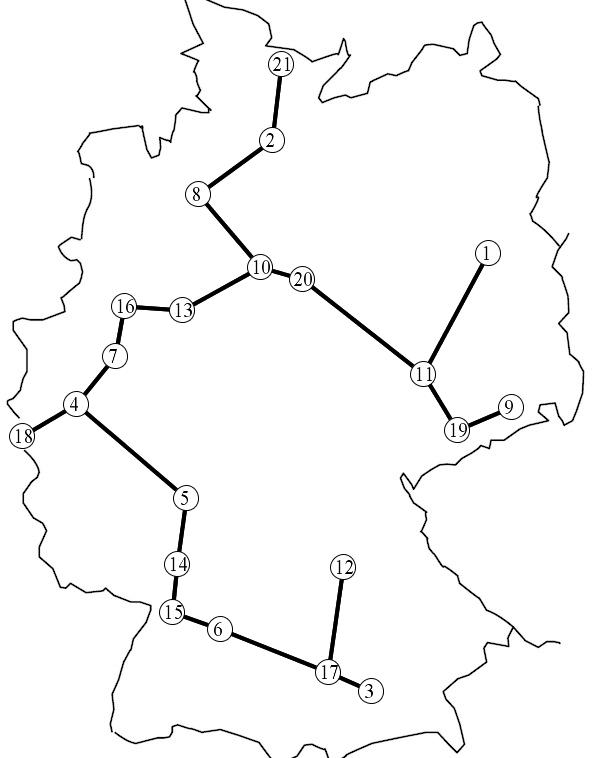}}
\subfigure[$\mathbf{ST}_5$]{\includegraphics[width=0.2\textwidth]{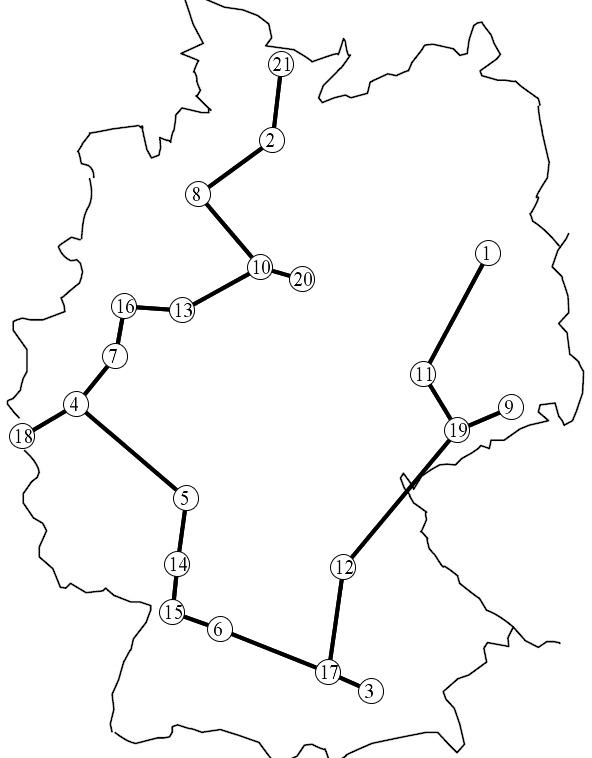}}
\subfigure[$\mathbf{ST}_{12}$]{\includegraphics[width=0.2\textwidth]{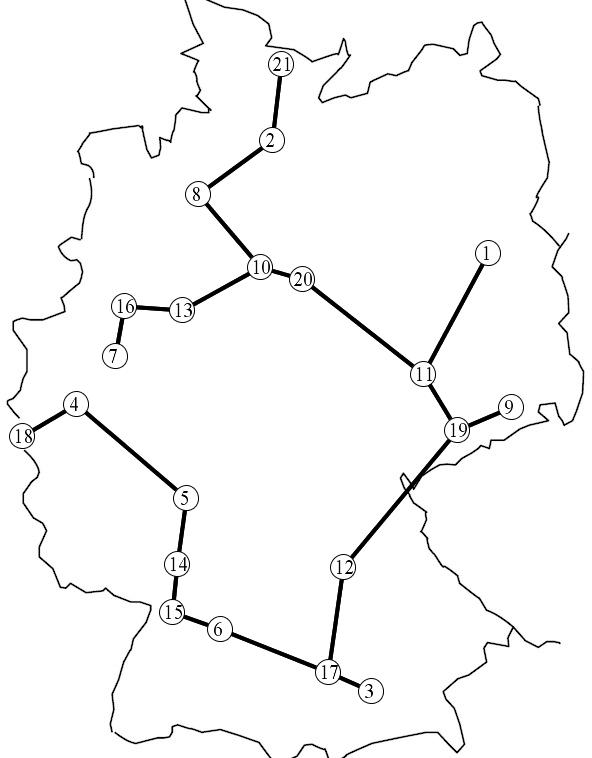}}
\subfigure[$\mathbf{ST}_{13}$]{\includegraphics[width=0.2\textwidth]{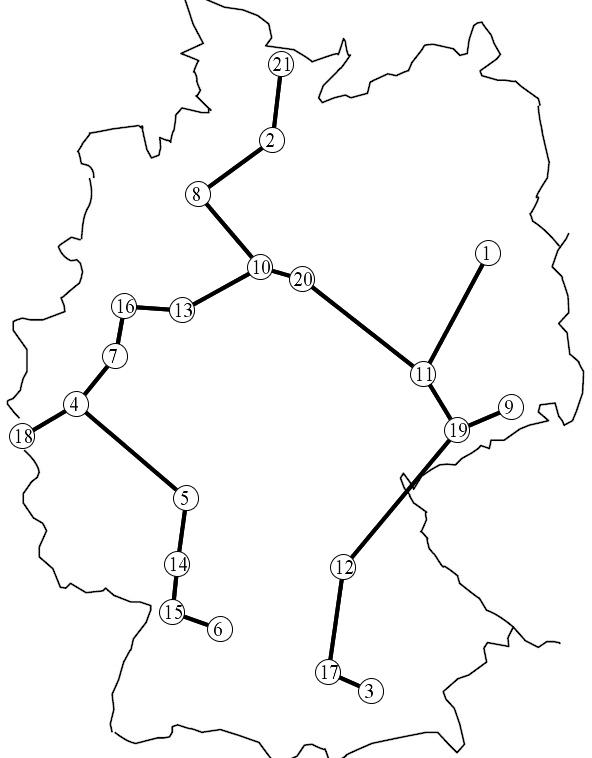}}
\caption{Topological classes of spanning trees, rooted in 
(a)~$\mathbf{ST}_1$: \One, \OneOne, \OneNine, \Nine, $l=1.02$.
(b)~$\mathbf{ST}_2$: \Two, \TwoOne, \Eight, \OneZero, $l=1.07$. 
(c)~$\mathbf{ST}_3$: \Three, \OneSeven, $l=1.07$.
(d)~$\mathbf{ST}_4$: \Four, \OneEight, \Seven, $l=1$.
(e)~$\mathbf{ST}_5$: \Five, \Six, \OneFour, \OneFive, $l=1.01$.
(f)~$\mathbf{ST}_{12}$: \OneTwo, $l=1.07$.
(g)~$\mathbf{ST}_{13}$: \OneThree, \OneSix, $l=1.04$.
Lengths $l$ of trees are normalised to a length of the minimum spanning tree (d).
}
\label{trees}
\end{figure}

The most common proximity graphs constructed on sites of $\mathbf{U}$ are shown in Fig.~\ref{proximity}. 
Seven topologies of spanning trees on $\mathbf{U}$ are shown in Fig.~\ref{trees}. Tree rooted in \One is not exactly the minimum tree but only 1.02 times longer.  

\begin{figure}[!tbp]
\centering
\subfigure[$\mathbf{H} \cap \mathbf{ST}_1$]{\includegraphics[width=0.2\textwidth]{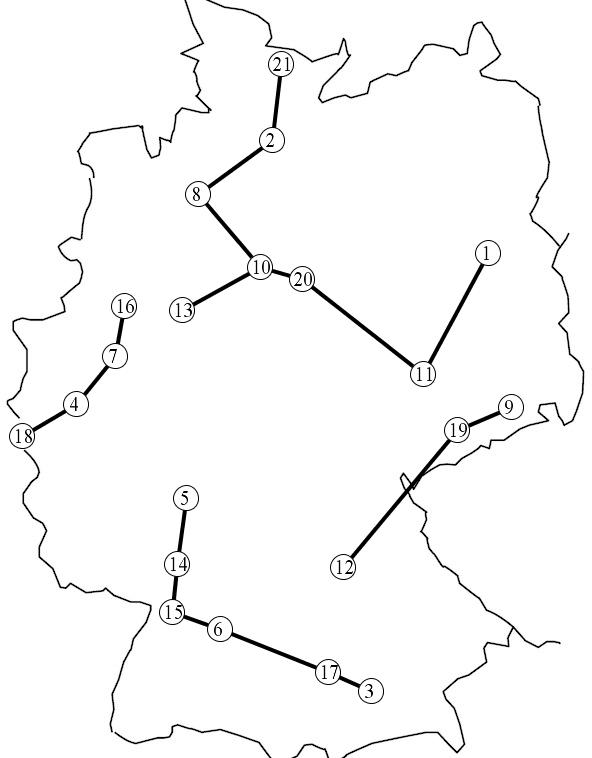}}
\subfigure[$\mathbf{H} \cap \mathbf{ST}_2$]{\includegraphics[width=0.2\textwidth]{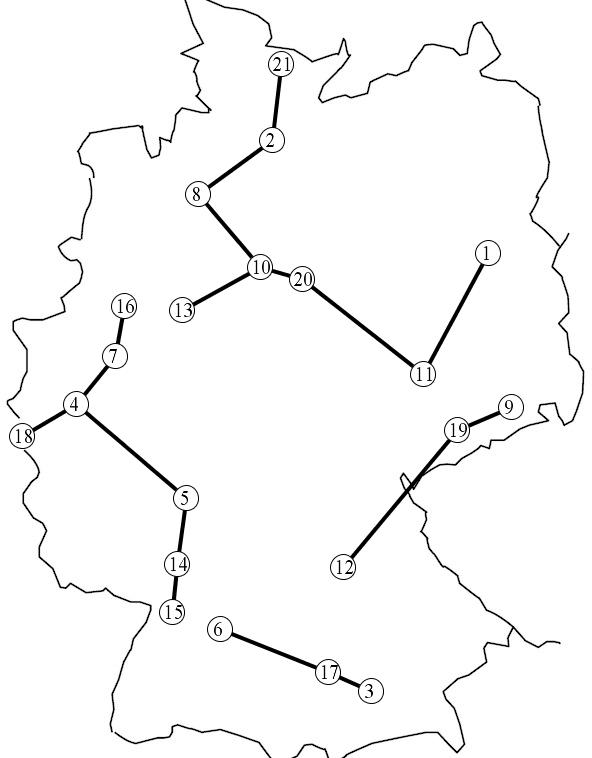}}
\subfigure[$\mathbf{H} \cap \mathbf{ST}_3$]{\includegraphics[width=0.2\textwidth]{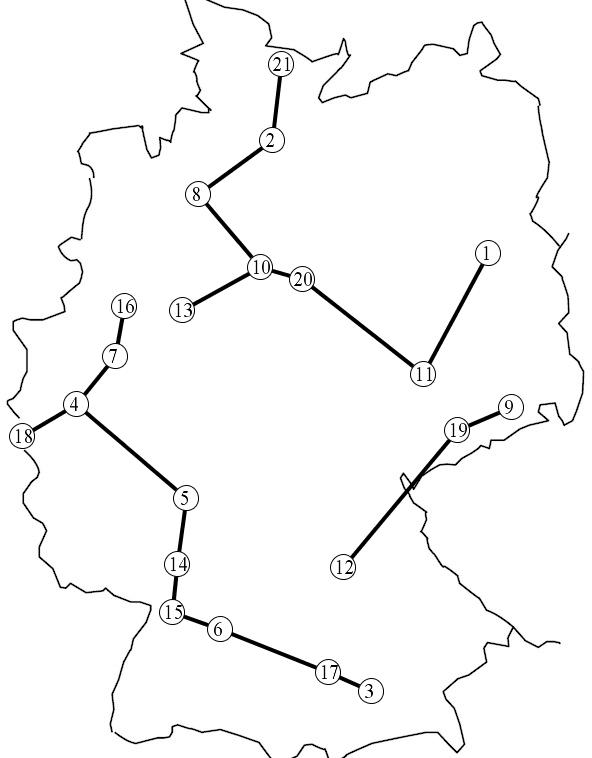}}
\subfigure[$\mathbf{H} \cap \mathbf{ST}_4$]{\includegraphics[width=0.2\textwidth]{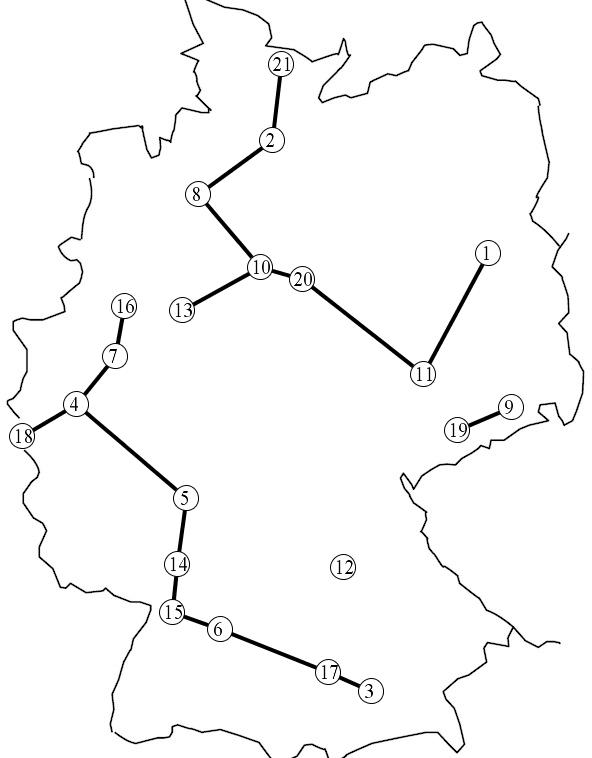}}
\subfigure[$\mathbf{H} \cap \mathbf{ST}_5$]{\includegraphics[width=0.2\textwidth]{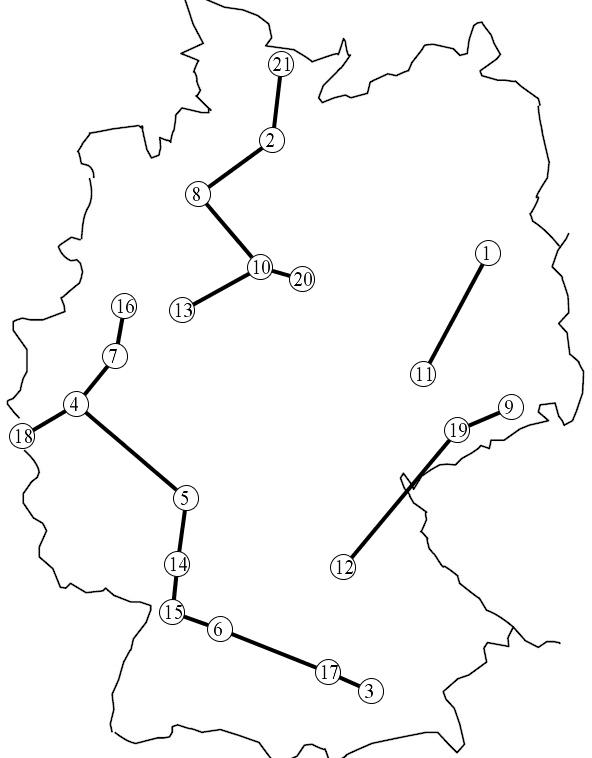}}
\subfigure[$\mathbf{H} \cap \mathbf{ST}_{12}$]{\includegraphics[width=0.2\textwidth]{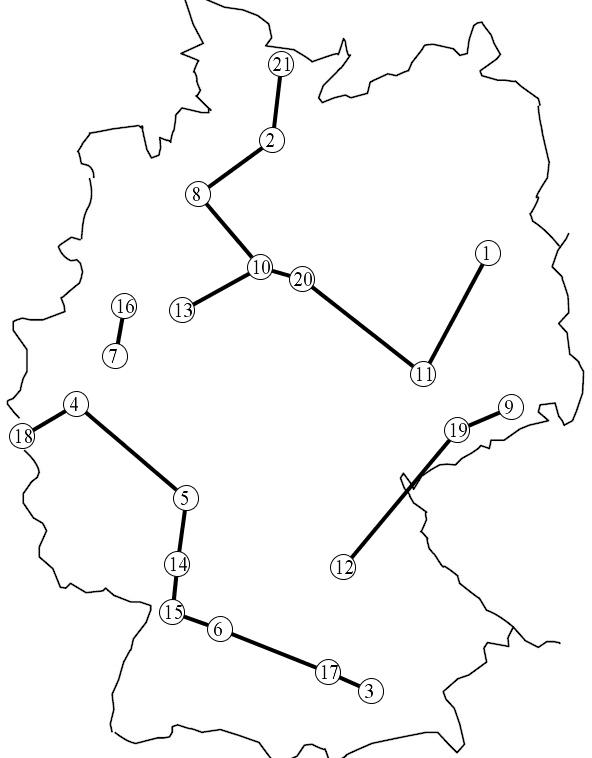}}
\subfigure[$\mathbf{H} \cap \mathbf{ST}_{13}$]{\includegraphics[width=0.2\textwidth]{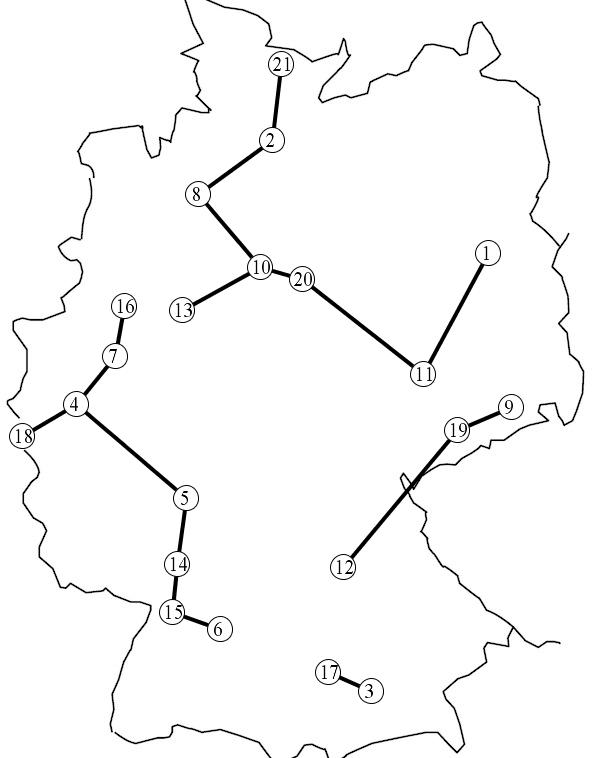}}
\caption{Intersection of  autobahn graph $\mathbf{H}$ with spanning trees 
(a)~$\mathbf{ST}_1$.
(b)~$\mathbf{ST}_2$.
(c)~$\mathbf{ST}_3$.
(d)~$\mathbf{ST}_4$.
(e)~$\mathbf{ST}_5$.
(f)~$\mathbf{ST}_{12}$.
(ag)~$\mathbf{ST}_{13}$.
}
\label{intersetmwaytrees}
\end{figure}

\begin{finding}
Let $E(\mathbf{G})$ be a number of edges in a graph $\mathbf{G}$ and 
$\mathbf{ST}$ be a spanning tree rooted from any node of $\mathbf{U}$ then 
$E(\mathbf{H} \cap \mathbf{ST}) = E(\mathbf{ST}) - 3$. 
\end{finding}

See illustration Fig.~\ref{intersetmwaytrees}.

\begin{figure}[!tbp]
\centering
\subfigure[$\mathbf{H} \cap {\mathbf{GG}}$]{\includegraphics[width=0.32\textwidth]{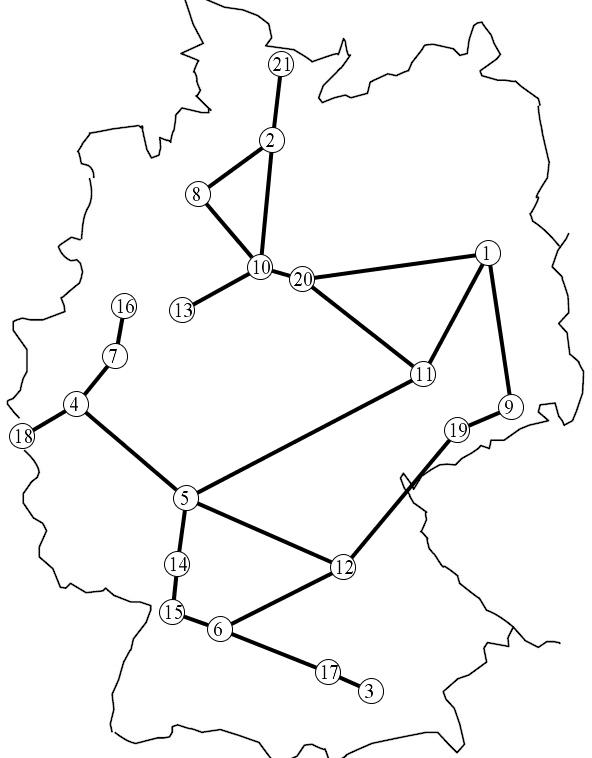}}
\subfigure[$\mathbf{H} \cap {\mathbf{RNG}}$]{\includegraphics[width=0.32\textwidth]{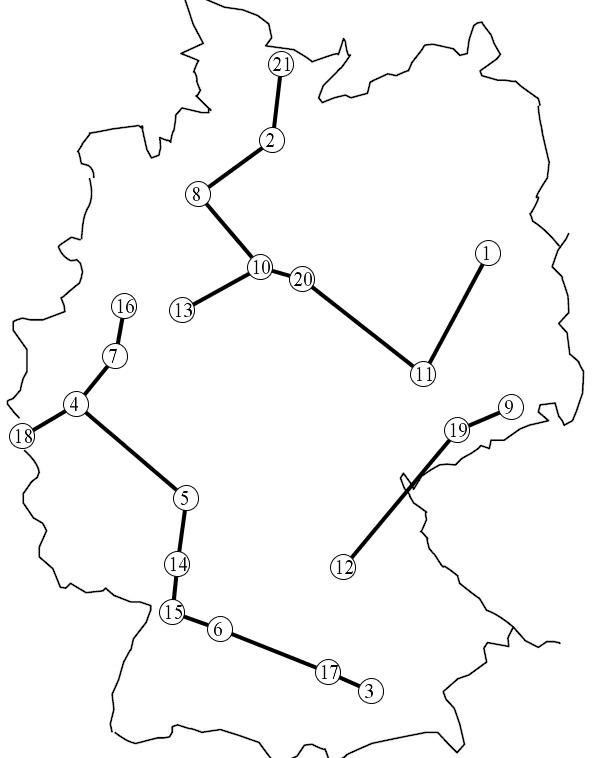}}
\caption{Intersection of autobahn graph with (a)~Gabriel graph and (b)~Relative neighbourhood graph.}
\label{intersetmwaysproximity}
\end{figure}

\begin{finding}
If $\mathbf{RNG}$ was a subgraph of $\mathbf{H}$ if $\mathbf{H}$ would have edges
(\OneThree -- \OneSix) and  (\OneTwo -- \OneSeven)  and 
$\mathbf{RNG}$ would have an edge (\OneOne -- \OneNine).
\end{finding}

\begin{finding}
$\mathbf{H} \cap \mathbf{RNG} = \mathbf{H} \cap \mathbf{ST}_{12}$
\end{finding}

Relative neighborhood graph $\mathbf{RNG}$, Gabriel graph $\mathbf{GG}$ and spanning trees $\mathbf{ST}$ are well known species of planar proximity graphs used in geographical variational analysis~\cite{gabriel_sokal_1969,matula_sokal_1984}, 
simulation of epidemics~\cite{toroczkai_2008}, and design of \emph{ad hoc} wireless networks~\cite{li_2004,song_2004,santi_2005,muhammad_2007,wan_2007}. The proximity graphs, particularly $\mathbf{RNG}$, are invaluable in simulation of human-made, road networks; these graphs are validated in specially interesting studies of Tsukuba central district road networks~\cite{watanabe_2005,watanabe_2008}. The graphs provide a good formal representation of biological transport networks, particularly foraging trails of ants~\cite{adamatzky_2002}. The fact that just two edges (\OneThree -- \OneSix) and  (\OneTwo -- \OneSeven) must be removed from $\mathbf{RNG}$ and only one edge removed from $\mathbf{RNG}$ to make the relative neighbourhood graph a sub-graph of the autobahn graph  demonstrates that the autobahn graph fits well into existing concepts of near optimal planar proximity graphs.

\begin{finding}
$\mathbf{ST}(a) \subset \mathbf{P}(\frac{1}{22})$ for any $a \in \mathbf{U}$.
\end{finding}

Namely, in laboratory experiments protoplasmic networks almost always maintain a minimum spanning core as an 
underlying structure of their topology. This can be demonstrated by direct comparison of graphs presented in 
Figs.~\ref{physarumgraphs} and \ref{trees}.

\begin{figure}[!tbp]
\centering
\subfigure[$\mathbf{RNG} \cap \mathbf{P}(\frac{1}{22})$]{\includegraphics[width=0.32\textwidth]
{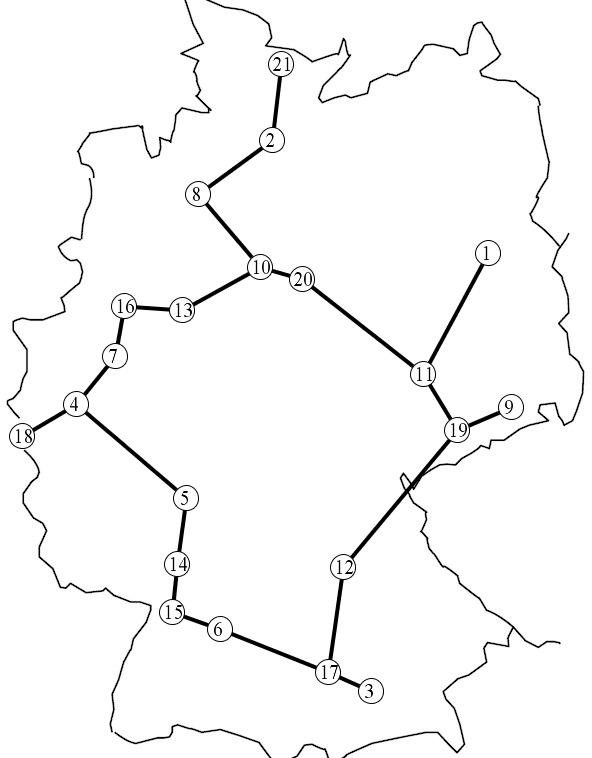}}
\subfigure[$\mathbf{GG} \cap \mathbf{P}(\frac{1}{22})$]{\includegraphics[width=0.32\textwidth] 
{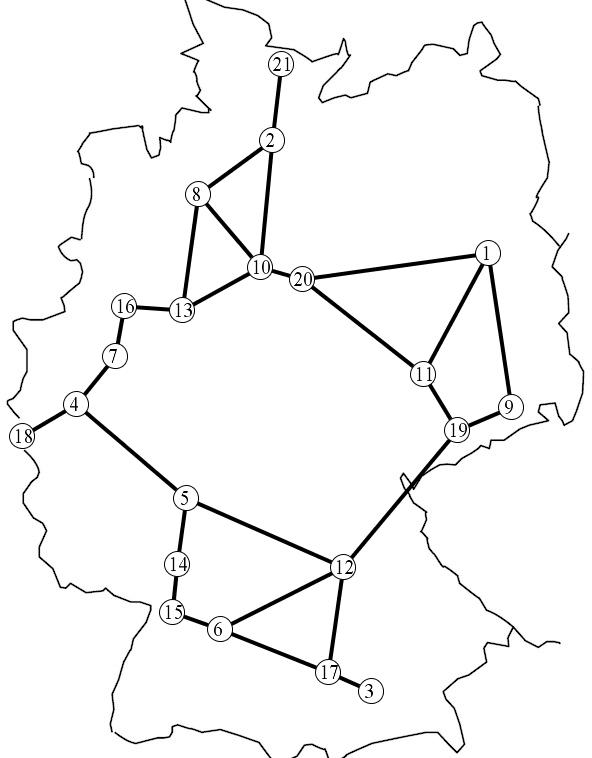}} \\
\subfigure[$\mathbf{RNG} \cap \mathbf{P}(\frac{7}{22})$]{\includegraphics[width=0.32\textwidth]
{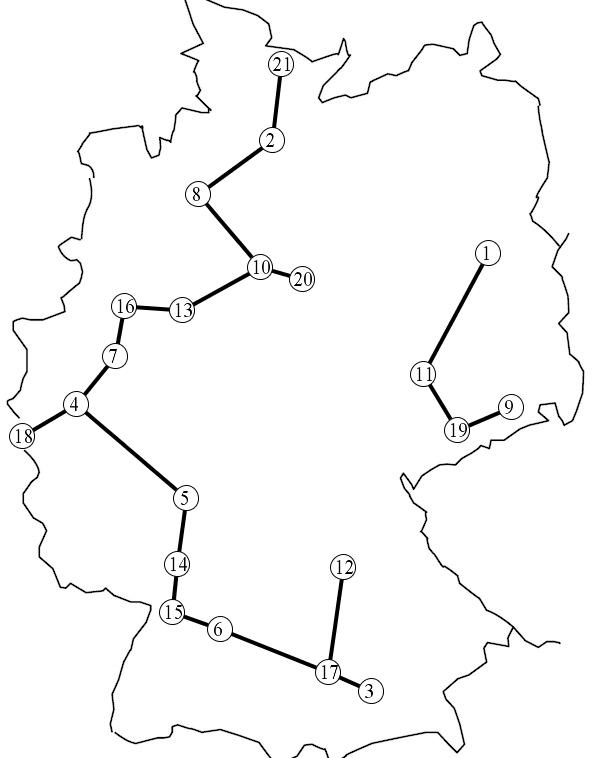}}
\subfigure[$\mathbf{GG} \cap \mathbf{P}(\frac{7}{22})$]{\includegraphics[width=0.32\textwidth]
{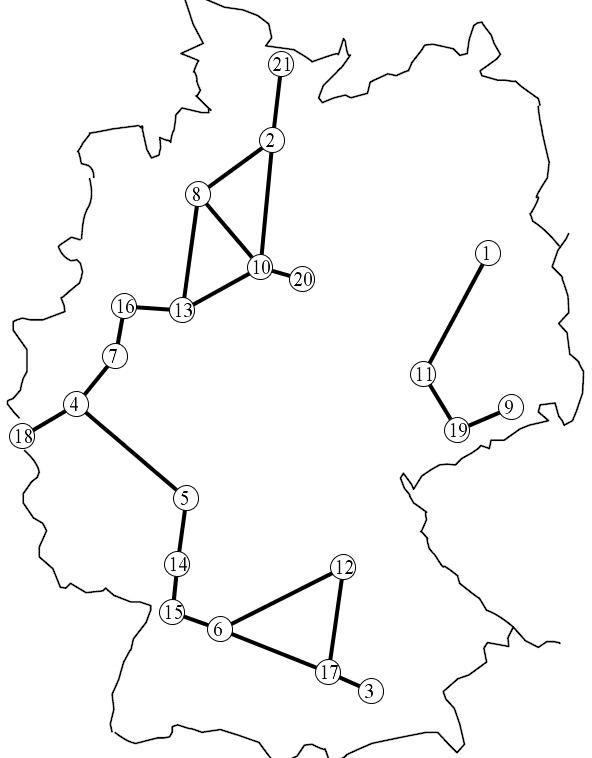}}
\caption{Intersection of Physarum graphs  (ab)~$\mathbf{P}(\frac{1}{22})$ and 
(cd)~$\mathbf{P}(\frac{7}{22})$ with  (ac)~relative neighbourhood graph and  (bd)~Gabriel graph.}
\label{proximitygraphsPhysarum}
\end{figure}

\begin{finding}
$\mathbf{GG} \subset \mathbf{P}(\frac{1}{22})$ 
\end{finding}

See Figs.~\ref{proximitygraphsPhysarum}. Gabriel graph is a super-graph of relative neighbourhood 
graph and of minimum spanning tree~\cite{toussaint_1980,matula_sokal_1984,jaromczyk_toussaint_1992}. Thus 
the slime mould \emph{P. polycephalum} imitates --- in its foraging patterns --- all three basic planar proximity 
graphs.
  
\clearpage

\section{Imitating large-scale contamination}
\label{disaster}

\begin{figure}[!tbp]
\centering
\subfigure[]{\includegraphics[scale=0.34]{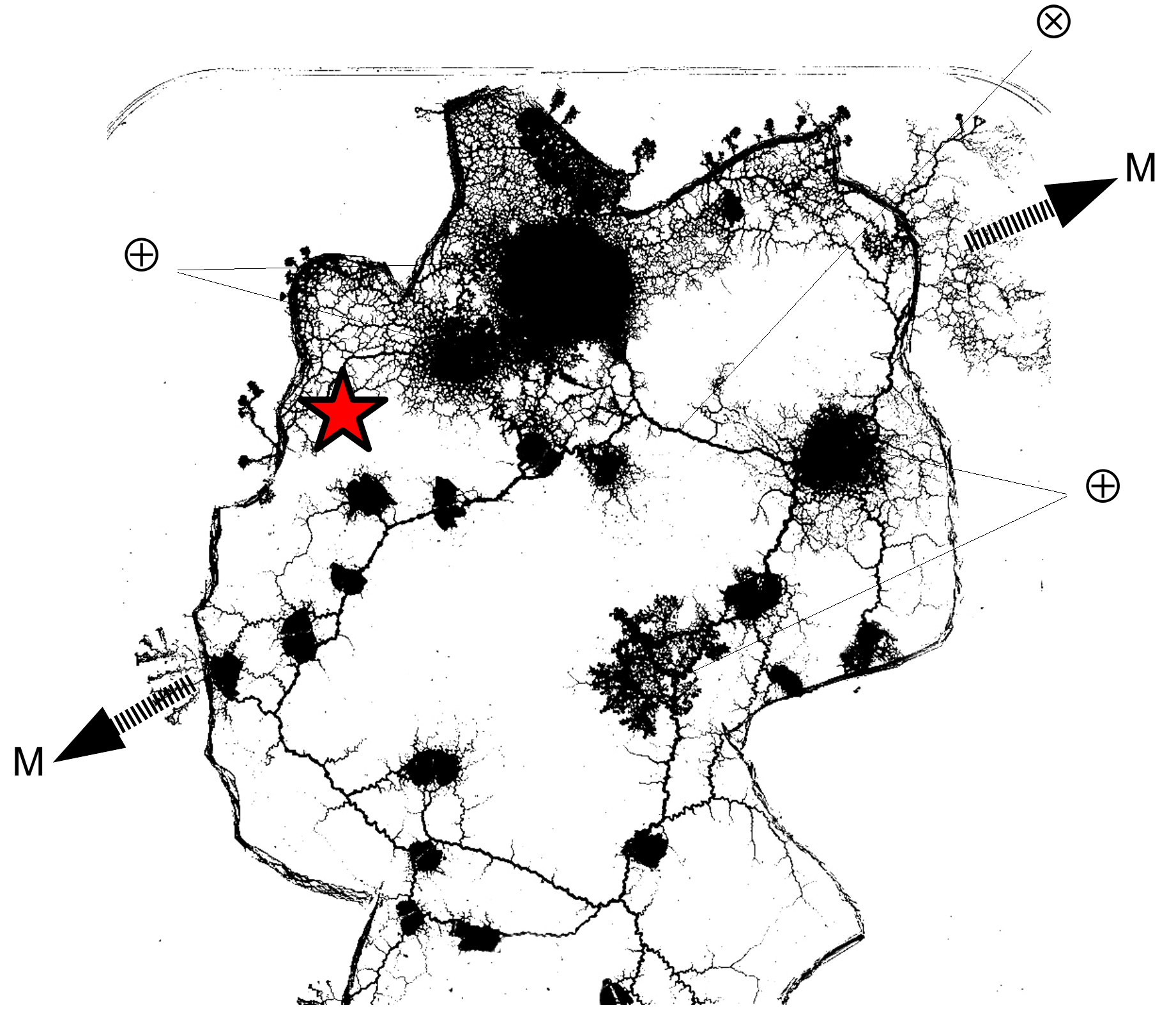}}
\subfigure[]{\includegraphics[scale=0.34]{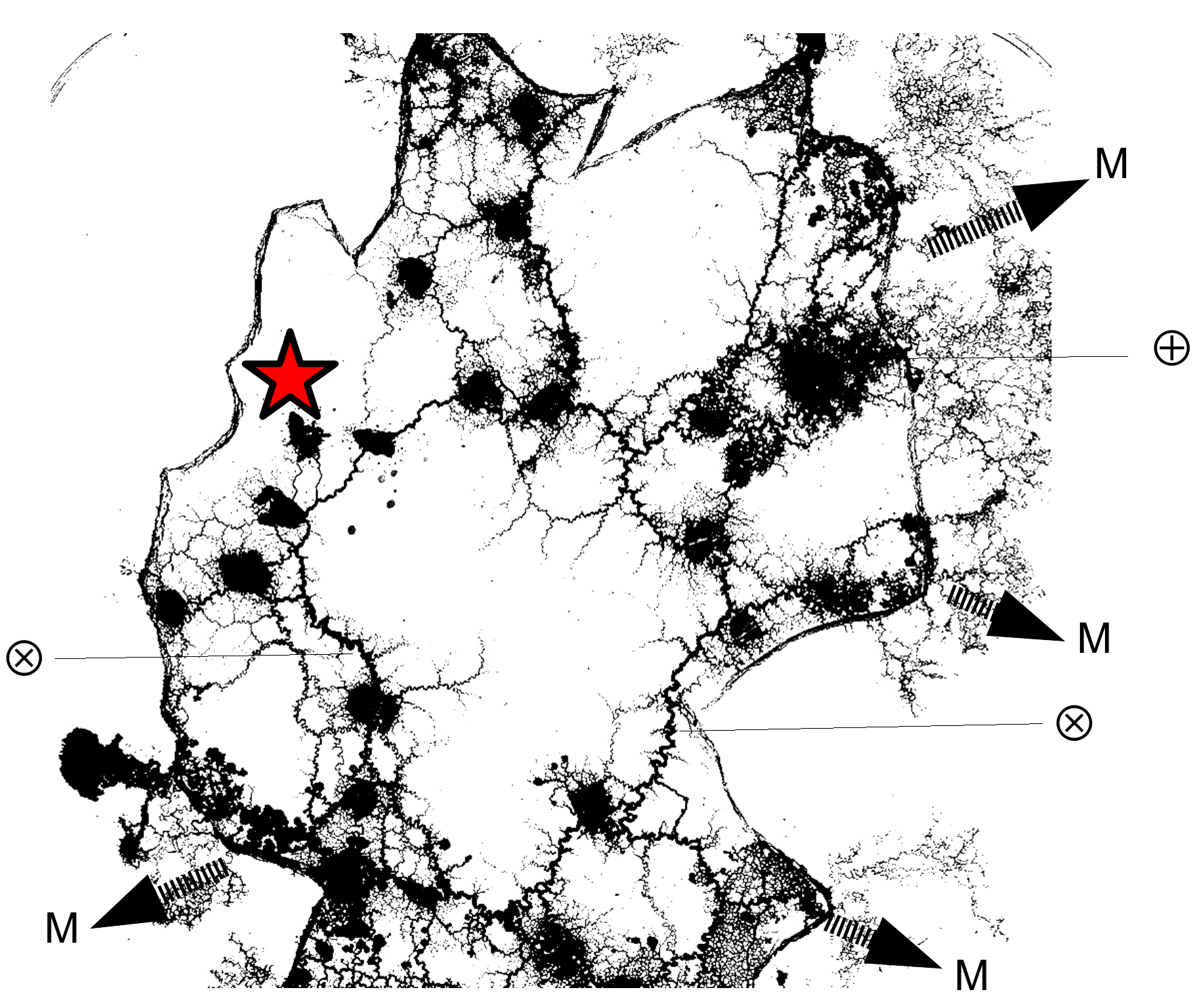}}
\subfigure[]{\includegraphics[scale=0.34]{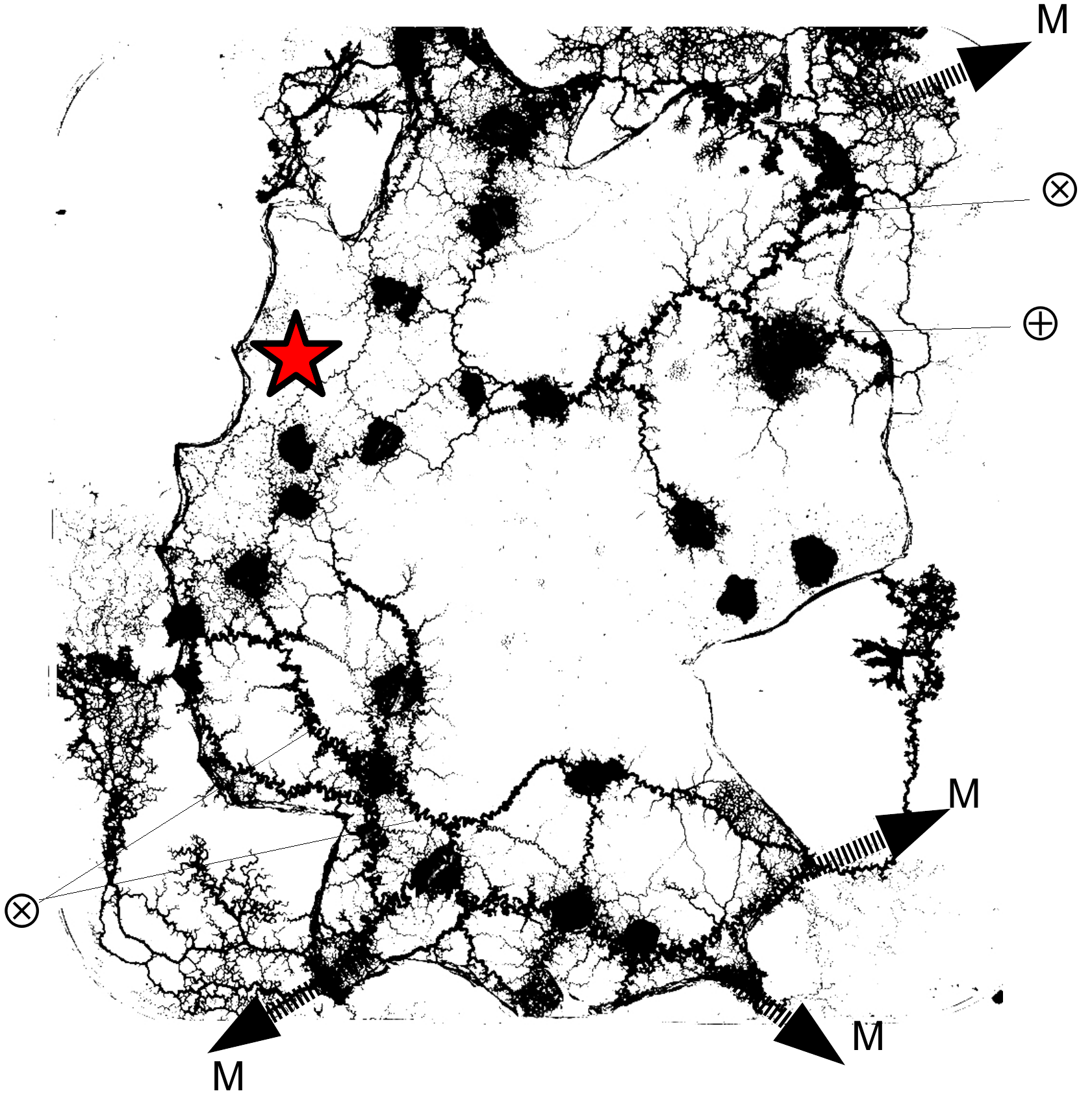}}
\caption{Exemplar snapshots of laboratory experiments on reconfiguration of protoplasmic network in a response to 
propagating contamination. The snapshots are taken 24~h after initiation of contamination. Mass-escape routes are marked by 'M', increase of activity in certain urban areas is labelled by $\oplus$ and increase in traffic along certain routes by $\otimes$. }
\label{salt}
\end{figure}

To study the effect of a large-scale contamination on a dynamic and structure of transport network, imitated by 
\emph{P. polycephalum} protoplasmic networks, we placed a grain of sea salt (SAXA Coarse Sea Salt, a crystal weight around 
20~mg) in the site of agar plate corresponding to the approximate position of Emsland Nuclear Power Plant~\cite{Emsland}.
The position is marked by a star in Fig.~\ref{salt}. Inorganic salt is a repellent for  \emph{P. polycephalum} therefore  the  sodium chloride diffusing into the plasmodium's growth substrate causes the plasmodium to abandon domains with high level of salinity. As an immediate response to diffusing chloride the plasmodium withdraws from the zone immediate to the epicentre of contamination.  Further response usually fits into two types: hyper-activation of transport as an attempt to deal with the situation and migration away from contamination, sometimes even beyond the agar plate, 
as an attempt to completely avoid the contaminated area. 

Typical scenarios of slime mould's response to contamination are illustrated in Fig.~\ref{salt}. Thus, we observe a substantial increase of foraging activity in Schleswig-Holstein, around Berlin and at the boundary between 
Thuringia and Upper Franconia. 

In example Fig.~\ref{salt}a we witness transport links are substantially enhanced between \One and \Two, \TwoZero and \OneZero and between \One and \OneOne and \Nine. Also, auxiliary transport link from \Two and \TwoOne to \One emerges along north-east boundary of the country (Fig.~\ref{salt}a); this link may play a key role in preparation of mass-migration from Germany to north-west Poland. Hyper-activity of transport links connecting \OneNine and \Three,
\OneTwo and \Six, \OneFive and \OneFour, and \Four and \OneFour is recorded in examples shown in 
 (Fig.~\ref{salt}bc).
  
Mass migration is observed from \OneEight to Limburg in Belgium, and from Mecklenburg, Western Pomerania inwards north-west Poland (Fig.~\ref{salt}a), from Dessau, Leipzig and Chemnitz areas towards Wroclaw in 
Poland (Fig.~\ref{salt}b), and from Baden-W\"{u}rttemberg area inwards south-east France and Switzerland (Fig.~\ref{salt}bc).

\section{Discussion}

In our experimental laboratory research we represented major urban areas of Germany with oat flakes and 
inoculated  plasmodium of \emph{Physarum polycephalum} in Berlin. The plasmodium propagated from Berlin to nearby 
urban areas then to fertile urban areas close to already colonised urban areas. Eventually all urban areas were 
colonised by the plasmodium.  We conducted 22 identical experiments.  We found that autobahn and protoplasmic networks match each other satisfactory in many integral characteristics and topological indices, especially connectivity, average link lengths and Randi\'{c} index. With regards to exact matching between edges of autobahn and Physarum graphs, in  40\% of laboratory experiments almost 60\% of the autobahn segments are represented by the slime mould's protoplasmic tubes. 

We found that only autobahn links  (\Five -- \TwoZero) and (\Five -- \OneOne)  are never represented by protoplasmic tubes
of \emph{P. polycephalum} in laboratory experiments.  The following transport links are imitated by the slime mould in over 70\% of laboratory experiments:  (\TwoOne -- \Two -- \Eight -- \OneZero --
\OneThree  -- \OneSix  -- \Seven  -- \Four  -- \Five  -- \OneFour  -- \OneFive  -- \Six -- \OneSeven --
\Three), (\OneZero -- \TwoZero), (\Four -- \OneEight), and (\OneSeven -- \OneTwo). The transport links presented by the plasmodium network in over 90\% of laboratory experiments are 
(\Two -- \Eight), (\OneZero -- \OneThree), (\Four -- \Seven), (\OneFour -- \OneFive -- \OneSix), and  
(\OneSeven -- \Three).  The man-made autobahn links represented in over half of laboratory experiments are the 
chain (\TwoOne -- \Two -- \Eight -- \OneZero -- \OneThree) with the branch (\OneZero -- \TwoZero) and 
the chain (\OneSix -- \Seven -- \Four -- \Five -- \OneFour -- \OneFive -- \Six -- \OneSeven -- \Three)  with the 
branch (\Four -- \OneEight). 

We did not aim to give any conclusive answer of weather autobahns are mathematically optimal and environmentally friendly or not.  We attempted to find how 'good' are autobahns from slime moulds' point of view, and how the transport network in Germany would develop if it was developed by foraging principles and bio-mechanics of the slime mould from scratch. We demonstrated that the  \emph{P. polycephalum} approximates autobahn satisfactory and thus can be 
considered as a valuable and user-friendly experimental laboratory tool for imitation of man-made transport networks with amorphous living creature.

\end{document}